\documentclass[cits]{PoS}

\title{What do fractals learn us concerning the masses of fundamental particles, of hadrons, and of nuclei? Concerning also disintegration life-times?}

\ShortTitle{Use of fractals to study particles, hadrons and nuclei masses}

\author{\speaker{Boris Tatischeff}%
  
        Institut de Physique Nucleaire Orsay\\
$^{1)}$ Univ. Paris-Sud, IPNO, UMR-8608, Orsay, F-91405, France\\
$^{2)}$CNRS/IN2P3, Orsay, F-91405, France\\
        
        E-mail: \email{tati@ipno.in2p3.fr}}

\abstract{The hadron spectroscopy is studied through the use of fractals and discrete scale invariance (DSI) implying log-periodic corrections to continuous scaling.  The masses of  mesons and baryons, reported by the Particle Data Group (PDG),  agree with (DSI), as well as the masses of exotic narrow mesons, baryons, and dibaryons. 

Two distributions are systematically studied: first the log of the masses versus the log of their rank,  and also the successive mass ratios. Each fitted parameter of the second distributions, as a function of the hadronic masses, displays the same shape for all PDG hadronic families and species.
The same parameters allow good fits for the narrow exotic mesons, baryons and dibaryons.

 When the successive mass ratios
between different baryon families are constant, this property is not observed between different meson families. Such observation is studied within the double mass ratios eliminating the quark masses, but the difference between baryons and mesons is not understood. 

The fractal properties and discrete scale invariance model are also used to study nuclei yrast masses as well as excited nuclei level masses of some nuclei. Here also
the good agreement between data and fractal property, allows to make some predictions for still unobserved nuclei masses.

Fractal properties are also compared to several nuclei data such as :\\
 - atomic masses in several columns of the Mendeleev periodic table of elements,\\
 - masses of series following $\beta^{+}$ or $\beta^{-}$ disintegrations,\\
 - one and two nucleon separation energies,\\
 - half-lives of some isotopes,\\
  - the four radioactive family periods.\\

Finally, it is shown that  the lepton, hadron, and boson masses can be presented in the same frame. This is also partially true for the coupling constants.}

\FullConference{XXI International Baldin Seminar on High Energy Physics Problems \\
		 September 10-15, 2012\\
		 JINR, Dubna, Russia
		 
PoS.cls\\
PoSlogo\\}

\begin{document}
		
\section{Introduction}
It is widely known, that fractal properties are very often observed in different physical fields like astrophysics, geoscience,  biology, paleontology, and others, and also in human sciences like history, economy, finance, and others \cite{nottale}. The large number of studies prevents any meaningfull quotation.   However  the broad field of applications is illustrated by two recent papers, devoted  to the study of Russian and Roman territories evolution \cite{ivan}, and to the Fractality and migrations of Homo sapiens \cite{ivan1}. 

Several models of hadronic and nuclei structures exist, emphasizing their single-particle or  collective natures. Beside these theoretical approaches, some  works (sometimes simple)  were devoted to the study of hadron properties concerning mainly masses. Many of them are reported here.

Terazawa explains and predicts fundamental particle masses in the unified composite model of quarks and leptons \cite{terazawa}. Here the minimal supersymmetric composite model of quarks and leptons consists of an isodoublet of spinor subquarks with charges $\pm$1.2, {\it w$_{1}$} and {\it w$_{2}$} (named "wakems" standing for weak and electromagnetic), and a Pati-Salam color-quartet of scalar subquarks with charges +1/2 and -1/6, C$_{0}$ and C$_{\it i}$ ({\it i = 1,2,3}) (named "chroms" standing for colors). 

Fu-Guang Cao determines \cite{cao} the absolute masses of neutrinos  by means of an intrinsic mass relation between leptons and quarks. Using the neutrino oscillation data, the following neutrino masses were obtained:  m$_1$ = 0.21+1.7-0.21 $10^{-4}$ eV,  m$_2$ = (8.7$\pm$0.1) 10$^{-3}$ eV, and m$_3$ = (4.9$\pm$0.1) 10$^{-2}$ eV.  The Koide relation for lepton masses was generalized by Kartavtsev \cite{kartavtsev} to the quark sector.

Several lepton and hadron masses were calculated by M$\ddot{\rm{u}}$eller \cite{mueller} and several other authors, using the "Continued fraction representations". Many calculated precise masses were obtained with 
use of some parameters. A recent paper by Ries reviews these results \cite{ries}. 

The quark-lepton similarity was studied by Hwang and Siyeon \cite{hwang}, and some relations between hadron masses were shown by Jacobsen \cite{jacobsen}.

The fractal theory applies to discontinuous evolutions with discret jumps, therefore it is precisely the theory to be applied to the masses of fundamental particles, hadrons and nuclei. 
\section{Recall of fractal properties}
Fractals exist in many aspects of nature. The characteristic of a fractal is that the relative value of an observable function of a variable, depends only on one parameter, and keeps the same form for different  variable values  \cite{mandelbrot} \cite{nottale1}. 

(a) The {\it continuous} scale invariance is defined in the following way: an observable O(x), depending on the variable x,  is scale invariant under the arbitrary change x $\to~\lambda$x, if there is a number $\mu(\lambda)$ such that O(x) = $\mu$O($\lambda$x). $\lambda$ is the fundamental scaling ratio. The solution of  O(x) is the power law:

\vspace*{-2.mm}
\begin{equation}
O(x) = C  x ^ { \alpha},  \hspace*{4.mm}   \alpha = -ln\mu/ln\lambda.
\end{equation}

(b) Unlike the {\it continuous} scale invariance, the {\it discrete} scale invariance (DSI) is observed when only specific $\lambda$ values satisfy (2.1). Then $\alpha$, the exponent of the power law is complex, inducing log-periodic corrections to the scaling \cite{sornette}. The $\alpha$ exponent is now:\\
\vspace*{-0.4cm}
\begin{equation} 
\alpha = -ln\mu/ln~\lambda + i 2n\pi/ln\lambda 
\end{equation}
where n is an arbitrary integer.

The recent observation of several relations connecting between them elementary particle masses \cite{btib}, suggests to look for fractal properties between these masses. The relations connect the first quark family masses with charge = 2/3, the second quark family masses with charge = -1/3, the gauge boson masses, and the lepton masses. The studies presented here  originate from such observation.

 Several hadronic and nuclear data, mainly masses, will be compared to a few linearities between ln(O) and ln(x), with different $\alpha$ values in the log-log distribution:
\vspace*{-0.1cm}
\begin{equation} 
 ln(O(x)) = \alpha~ln(x) + ln(C)
 \end{equation} 
 \vspace*{-0.6cm}
 
  The following figures compare the data to the previous property, namely the possible straight lines in the ln(M) = f(ln(rank)), where M is the mass of a given species, and  the rank denotes the successive values of the masses.

The most general form of a distribution is \cite{sornette}:
\vspace*{-0.2cm}
\begin{equation}
f(r) = C\hspace*{1.mm}(|r-r_{c}|)^{s}\hspace*{1.mm}[1 + a_{1}\hspace*{1.mm}cos(2\pi\hspace*{1.mm}\Omega\hspace*{1.mm}
ln\hspace*{1.mm}(|r-r_{c}|) + \Psi)] 
\end{equation}

\vspace*{-0.2cm}

which is used to fit the successive mass ratios by adjusting the parameters. The comparison between the successive mass ratios and fitted curves will be the second test between data and fractal properties.

$a_{1}$ measures the amplitude of the log-periodic correction to continuous scaling. The critical exponent "s" is defined by $\mu$ = $\lambda^{s}$. $\Omega$ = 1/ln$\lambda$ measures the significance of the log-periodic corrections, therefore determines the $\alpha$ imaginary part.
The variation of the three main parameters of (2.4): "s", "a$_{1}$", and "$\Omega$"  will be discussed later. 

Other less important parameter is "$r_{c}$" the  critical rank, which describes the transition from one phase to another. It is underdetermined, but widely larger than the experimental "r" values for the more or less constant oscillation widths. This is the general situation of hadronic mass ratios.
Such undetermination has very few consequences on all parameters, except
on $\Omega$. When increasing "$r_{c}$" from 30 to 40, $\Omega$ increases approximately by a factor 1.36. Therefore "$r_{c}$"  was arbitrarily fixed to "$r_{c}$" = 40 for all following studies. Other less important parameters are $\Psi$, a phase in the cosine, and C a normalization constant.

\section{Application to some atomic data and quark masses}
The fractal relations reminded above were recently applied to the study of the fundamental coupling constants, atomic masses and energies, and elementary particle masses \cite{boris}.
\begin{figure}[ht]
\begin{center}
\scalebox{.9}[.5]{
\includegraphics[bb=24 237 522 550,clip,scale=0.7]{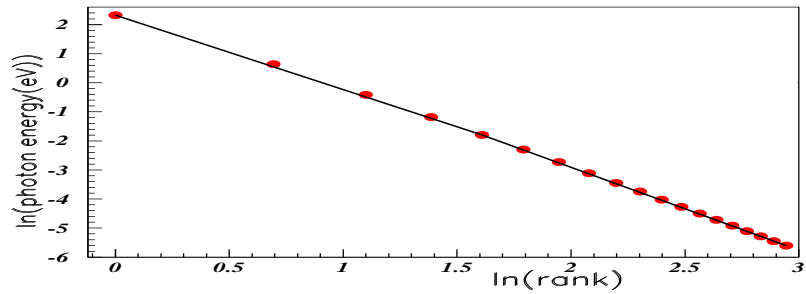}}
\caption{Photon energies in logarithmic scales between successive principal quantum numbers, versus logs. of the ranks.}
\end{center}
\end{figure}

The atomic level energies are given by the Rydberg formula:  $E_{n} = Z^{2}R_{E}/n^{2}$. We get obviously a linear relation between ln( $E_{n}$) and ln(n): ln($E_{n}$) = ln(C) - 2ln(n). $R_{E}$ is the Rydberg energy and "n" the principal quantum number.

The photon energies between two different principal quantum numbers ($n_{i}$) and ($n_{f}$) are given by the relation:  E = $R_{E}(1/n^{2}_{f} - 1/n^{2}_{i})$. Although the linear property is no more analytical, it is numerically still present as can be seen in fig.~1.
\begin{figure}[ht]
\begin{center}
\scalebox{.9}[.6]{
\includegraphics[bb=14 144 522 547 clip,scale=0.7]{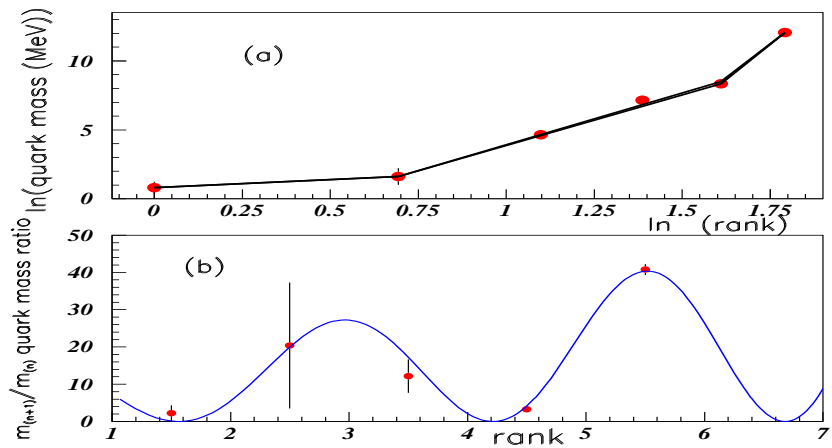}}
\caption{Insert (a) shows the log-log distribution of quark masses versus their rank. Insert (b) shows the quark successive mass ratio distribution.}
\end{center}
\end{figure}

The linearity between the logarithms of four quark masses is shown in fig.~2(a) and the successive mass ratio of the same masses is shown in fig.~2(b). The masses used are: m(u) = 2.28~MeV, m(d) = 5.1~MeV, m(s) = 101~MeV, m(c) = 1270~MeV, m(b) = 4200~MeV, and m(t) = 172,000~MeV. The discussion on the parameters, extracted from fits using the equation (2.4), will be presented below.

The bad knowledge of the neutrino masses prevents to look at the lepton fractal properties since we remain with only three known masses. The three boson masses W, Z, and H are aligned as can be seen at the end of the paper.
The common fractal property between quark, lepton and boson masses, and the coupling constants will be discussed at the end of the paper.

\section{Application to hadronic masses}

All masses are taken from the review of the Particle Data Group (PDG) \cite{pdg},  except those specifically indicated.
Many masses, omitted from the PDG summary table, but given in the detailed table contents, are kept in 
our study, which therefore will have to be improved after new mass determinations (observations, confirmations or eliminations). 

All baryonic and mesonic families including at least several known masses, have been studied {\cite{bor2} and compared to equations (2.3) and (2.4). A selection is presented here. 
\subsection{Mesons}
Fig.3 shows log-log plot and successive mass ratio distributions of two mesonic families.
\begin{figure}[ht] 
\begin{center}
\scalebox{1.3}[0.75]{
\includegraphics[bb=18 144 527 562,clip,scale=0.6]{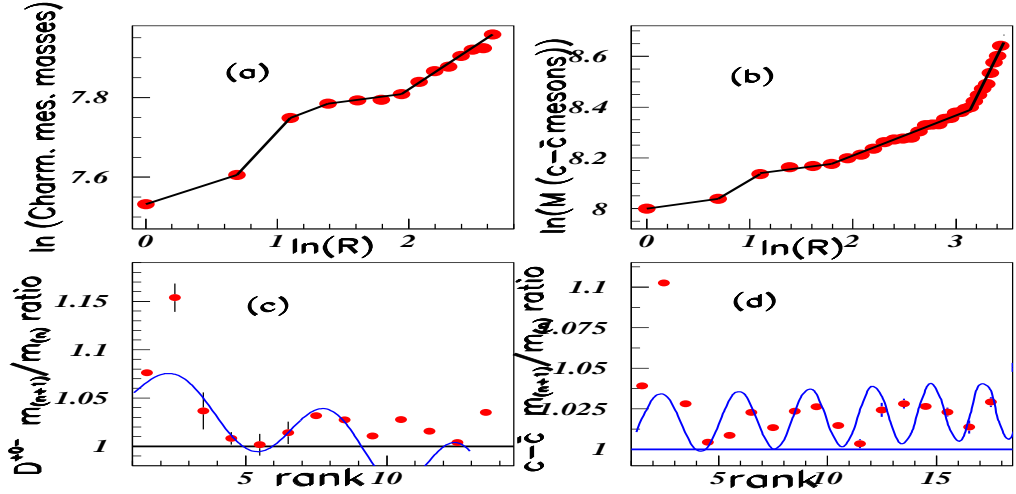}}
\caption{Log-log plot and successive mass ratio distributions of two mesonic families. Inserts (a) and (c) show the distributions corresponding to charmed mesons $D^{+}(c{\bar d})$,  $D^{0}(c{\bar u})$, and  $D^{-}(d{\bar c})$ in the range 1867$\le$M$\le$2860~MeV. Inserts (b) and (d) show the distributions of charmonium $c{\bar c}$ mesons in the range 2980$\le$M$\le$5910 MeV.} 
\end{center}
\end{figure}
\vspace*{-0.2cm}
\begin{figure}[ht]
\begin{center}
\scalebox{1.3}[0.75]{
\includegraphics[bb=14 137 524 548,clip,scale=0.6]{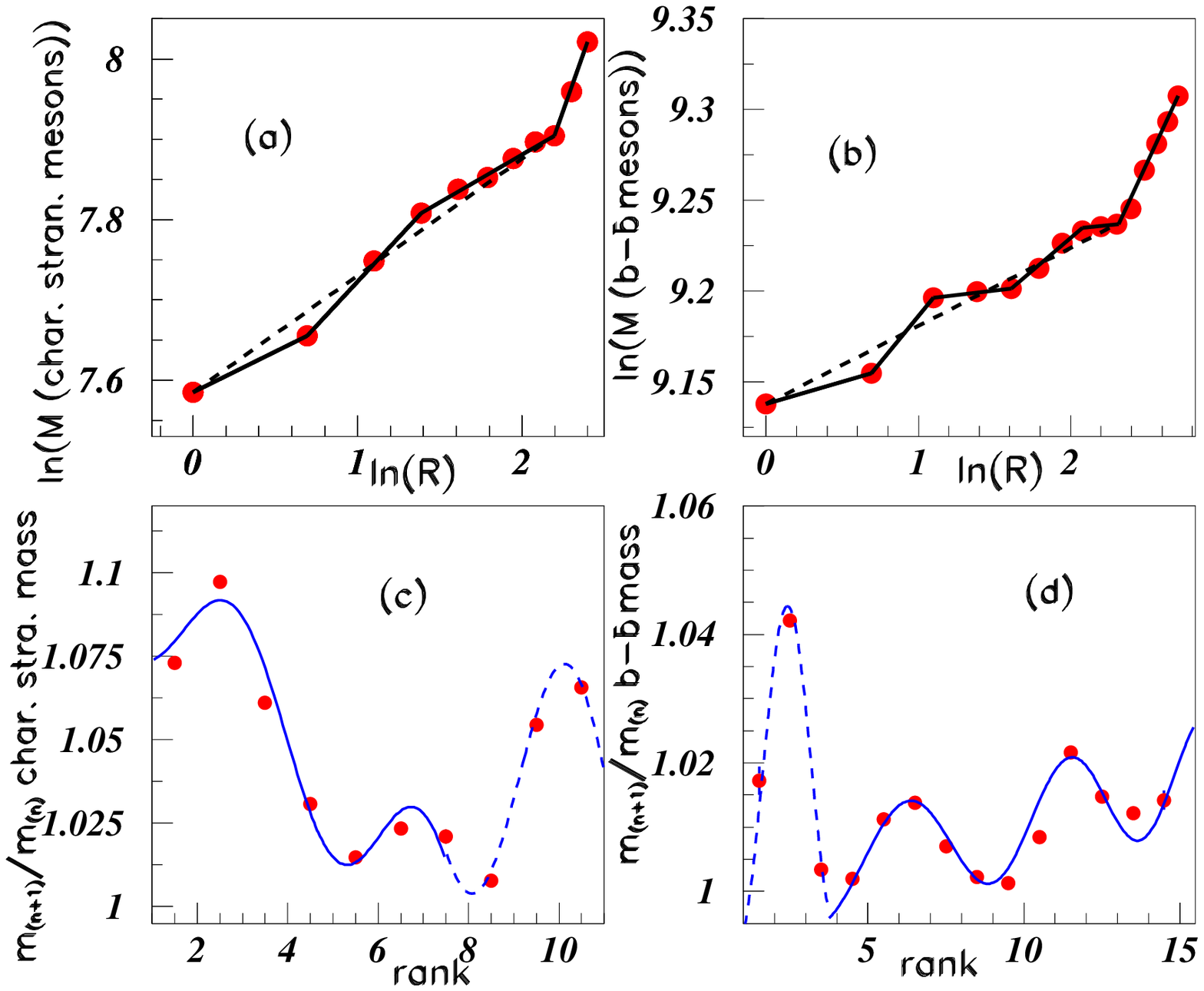}}
\caption{Log-log plot and successive mass ratio distributions of two mesonic families. Inserts (a) and (c) show the distributions corresponding to charmed strange mesons $D^{+}_{S}(c{\bar s})$ and  $D^{-}_{S}(s{\bar c})$ in the range 2862$\le$M$\le$3044~MeV. Inserts (b) and (d) show the distributions of bottonium $b{\bar b}$ mesons in the range 9300$\le$M$\le$11019 MeV.} 
\end{center}
\end{figure}
 Inserts (a) and (c) show the distributions corresponding to charmed mesons $D^{+}(c{\bar d})$,  $D^{0}(c{\bar u})$,  and  $D^{-}(d{\bar c})$ in the range 1867$\le$M$\le$2860~MeV. Inserts (b) and (d) show the distributions of charmonium $c{\bar c}$ mesons in the range 2980$\le$M$\le$5910 MeV. Inserts (a) and (b) show the log-log distributions. We observe straight line segments: signature of fractal presence. Inserts (c) and (d) show the successive mass ratio data, fitted with equation (2.4). The ratios between rank "n" and "n+1" are drawn at rank "n+1/2".
 With an unique parameter set, the data are reproduced from rank 3.5 up to rank $\approx$~10.

Fig.4 shows again the log-log plots and successive mass ratio distributions of two different mesonic families. Inserts (a) and (c) show the distributions corresponding to charmed strange mesons $D^{+}_{S}(c{\bar s})$ and  $D^{-}_{S}(s{\bar c})$ in the range 2862$\le$M$\le$3044~MeV. Inserts (b) and (d) show the distributions of bottonium $b{\bar b}$ mesons in the range 9300$\le$M$\le11019$ MeV . Here again inserts (a) and (b) show the log-log distributions and inserts (c) and (d) show the successive mass ratio data and fits. The same comments, as those concerning fig.~3, can be presented here, except that the fits inside inserts (c) and (d) are obtained with the use of two sets of parameters. Notice that the data in both inserts (a) and (b) can also be fitted with only two segments (dashed lines) but then a few data remain a little outside the straight lines.
\begin{figure}[ht]
\begin{center}
\scalebox{1.3}[0.75]{
\includegraphics[bb=14 135 524 550,clip,scale=0.6]{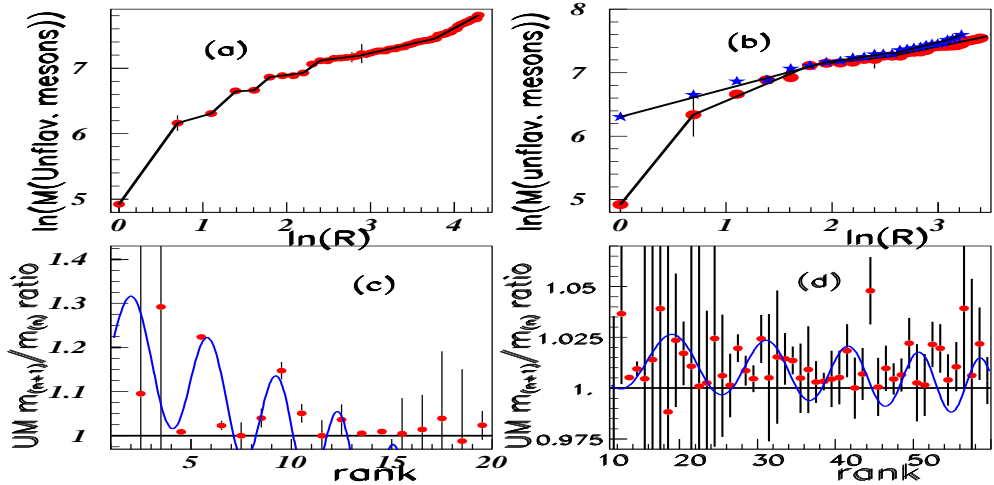}}
\caption{Log-log plot and successive mass ratio distributions of  the unflavoured mesons (see text).} 
\end{center}
\end{figure}

Fig.~5 shows the same analysis applied to the unflavoured mesons. Insert (a) shows the log-log distribution, which displays a clear staircase or jagged shape. We observe it  for the first six masses which suggests a possible double fractal property. Indeed, in the log-log representation, the three masses at R = 2, 4, and 6 are aligned in one hand, and the five masses at R = 3, 5, 7, 10, and 11 are also aligned in the other hand. 
We also observe that the product of the P parity by the G parity is successively even and odd for the first thirteen unflavoured meson masses, up to {\it f$_{2}$ (1270)}. We plot therefore for all unflavoured mesons two separeted log-log-distributions in figure 5(b). The even (odd) PG parity masses are considered separately. Full red circles are for even PG parity masses and full blue stars are for odd PG parity masses.
The staircase shape is then removed. 

 We deduce the presence of power-law sequences, for both even and odd PG parity products. These  relations could eventually help to predict the masses of unflavored mesons, built on the pion (or the $\eta$) mass. 

The $m_{n+1}/m_{n}$ unflavored mesonic mass ratios are plotted in inserts (c) and (d). 
 The first ratio is much larger than the other ratio values, and is removed from the figs.~5(c) and 5(d). It corresponds to the large mass difference between the two first unflavoured pions. The masses of two {\it $f_{0}$} mesons are very imprecise. We use 
 M($f_{0}(600)$) = 600$\pm$200~MeV, and M($f_{0}(1370)$) = 1370$\pm$200~MeV.
 
 The masses are introduced in an increasing order, therefore, in a few cases, the order is not exactly the same as the one given by PDG.  
 
Insert (c) shows the low rank ratios and fit performed using eq.~(2.4) with a single set of parameters. A good fit is observed up to rank 13.
 Insert (d) shows the ratio in a larger rank range. In this range, the good alignement of the log-log distribution, shown in insert (b), could suggest a better justification of the analytical fit with a single set of parameters. However the comparison between the data and the fit is
difficult because of the so large error bars. 
It is clear that this meson family contains a large number of very imprecise masses.
\subsection{Baryons}
Here we recall some results published in \cite{bor2}.
\begin{figure}[ht]
\begin{center}
\scalebox{1.3}[0.75]{
\includegraphics[bb=21 138 520 545,clip,scale=0.6]{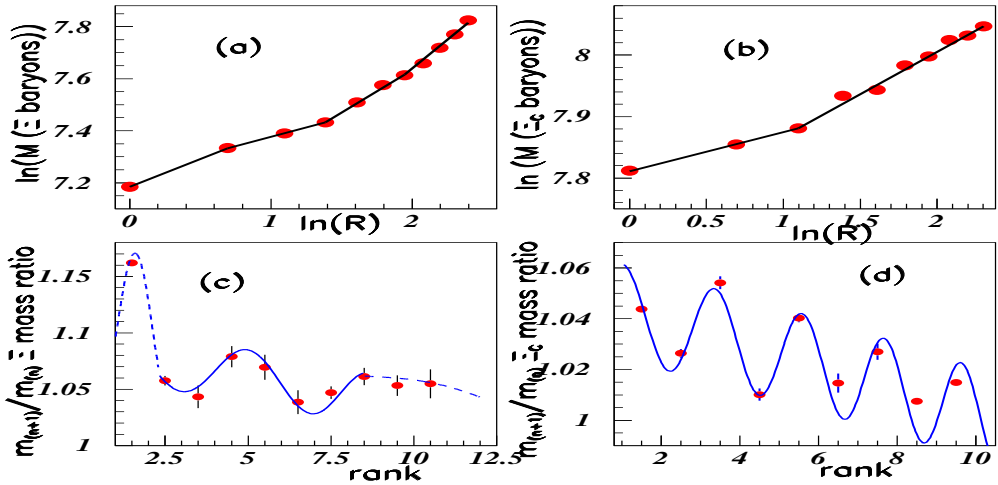}}
\caption{Log-log plot and successive mass ratio distributions of two baryonic families. Inserts (a) and (c) show the distributions corresponding to $\Xi$ (uss and dss) baryons, inserts (b) and (d) show the distributions corresponding to $\Xi_C$ (usc and dsc) baryons (see text).} 
\end{center}
\end{figure}

Fig.6 shows the results for the $\Xi$ baryons in inserts (a) and (c) and $\Xi_{C}$ baryons in inserts (b) and (d). Nice alignements are observed in the logarithms of the mass versus the logarithms of the rank distributions in figs. 6(a) and 6(b). The successive mass ratios are well fitted in inserts (c) (three sets of parameters) and (d) (a single set of parameters). The parameter variations will be discussed later.
\begin{figure}[ht]
\begin{center}
\scalebox{1.3}[0.75]{
\includegraphics[bb=21 138 530 550,clip,scale=0.6]{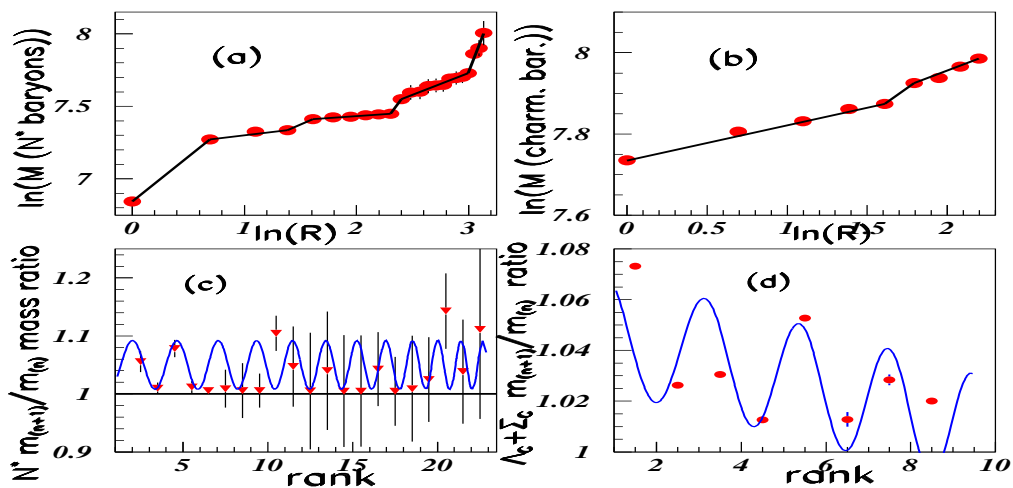}}
\caption{Log-log plot and successive mass ratio distributions of two baryonic families. Inserts (a) and (c) show the distributions corresponding to N$^{*}$ baryons, inserts (b) and (d) show the distributions corresponding to the charmed $\Lambda_{C}, \Sigma_{C}$ ( udc, uuc, ddc) baryons.} 
\end{center}
\end{figure}

Fig.7 shows the results for the N$^{*}$ baryons in inserts (a) and (c) and 
$\Lambda^{+}_{C}, \Sigma^{++}_{C}$ baryons in inserts (b) and (d). Again alignements are observed in the log of the mass versus the log of the rank distributions in inserts (a) and (b). The successive mass ratios are only fitted in fig. 7(c) up to rank 8.5 with a single set of parameters. If we except the first six masses, the greater N$^{*}$ masses are very imprecise.
Insert (d) shows the successive mass ratios of the charmed baryons (uuc, udc, ddc). An unique set of parameters allows to get a good fit to these data.
 The parameter variations will be discussed later.
\begin{figure}[ht]
\begin{center}
\scalebox{1.3}[1.0]{
\includegraphics[bb=37 134 520 520,clip,scale=0.6]{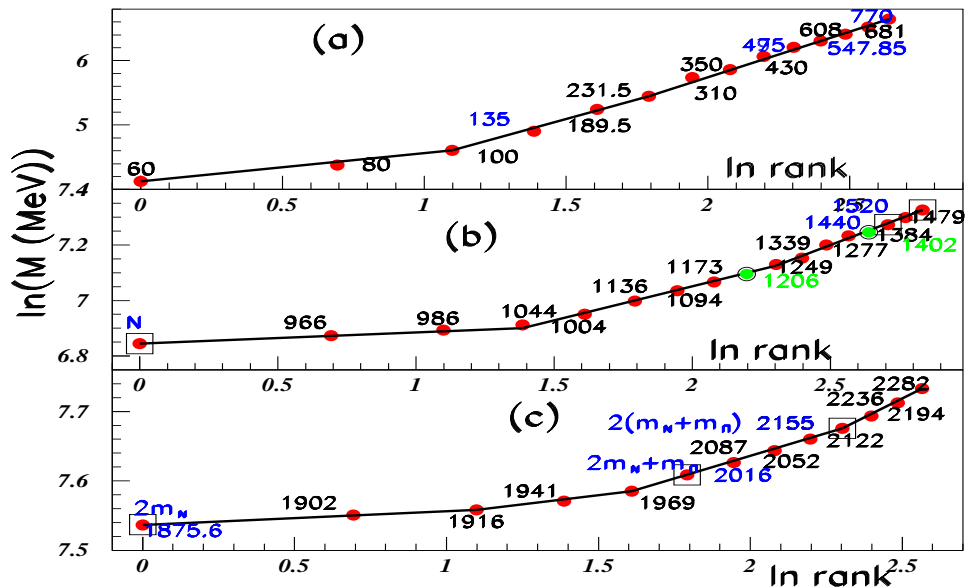}}
\caption{Log-log distributions of narrow exotic meson masses (insert (a)), narrow exotic baryon masses (insert (b)), and narrow exotic dibaryon masses (insert (c)) (see text).} 
\end{center}
\end{figure}
\subsection{Application to narrow baryons}
The previous data concerned hadronic masses from PDG \cite{pdg}.

 Some years ago, narrow and weakly excited structures were observed in  unflavoured mesons \cite{bormes},  baryons \cite{borbar}, and dibaryons \cite{bordibar}. Thanks to good resolution and high statistics, these weakly excited structures were observed at fixed masses, independent of the experimental device, reaction, incident energy, and spectrometer angle. Main data came from the SPES3 beam line at the Saturne synchrotron (Saclay Laboratory); they were observed in the missing mass and the invariant mass of the pp$\to$ppX and pp$\to$p$\pi^{+}$X reactions, studied at different incident energies and angles. Additionnal data came from Saturne Spes1 beam line, $^{3}$He(p,d)X reaction \cite{spes1} and Saturne Spes4 beam line, p($\alpha,\alpha^{'}$)X reaction \cite{spes4}. The agreement concerning the narrow exotic baryonic structure masses was obtained from data obtained at Spes3 and Spes4 although these reactions were studied for different motivations and published before the analyses performed to look for narrow structures. These results were obtained by different physicists, using different experimental set-ups, and different beam lines and reactions. 

Fig.~8 shows the log-log distributions of:\\
  -  narrow exotic mesons in insert (a)  \cite{bormes},\\
  - narrow exotic baryons in insert (b) \cite{borbar},\\
  - narrow exotic dibaryons in insert (c)  \cite{bordibar}. 
 
 Some very  small mesonic structures close to M $\approx$ 700~MeV have been removed, since they were not observed in the measurement performed at JINR (Dubna) \cite{troyan1}. In the same way, the very small baryonic structure at M $\approx$ 953~MeV has been removed. The two baryonic encircled green data at M = 1206 and 1402~MeV, are introduced although they were not extracted from the Spes3 measurements. The extraction of a small structure at M = 1206~MeV  was difficult since it lies close to a large pic corresponding to $\Delta$(3,3).  A structure at M = 1402~MeV  could only be observed in the missing mass at the largest incident proton energy ($T_{p}$ = 2.1~GeV) and the two smallest spectrometer angles: $\theta$ = 0.7$^{0}$ and 3$^{0}$. Both missing mass spectra are again rather large in this mass range, due to phase space background and contributions from broad resonances.
 
 We observe again straight line segments in all three inserts (a), (b), and (c) of fig.~8.
\begin{figure}[ht]
\begin{center}
\scalebox{1.2}[0.75]{
\includegraphics[bb=7 101 520 553,clip,scale=0.6]{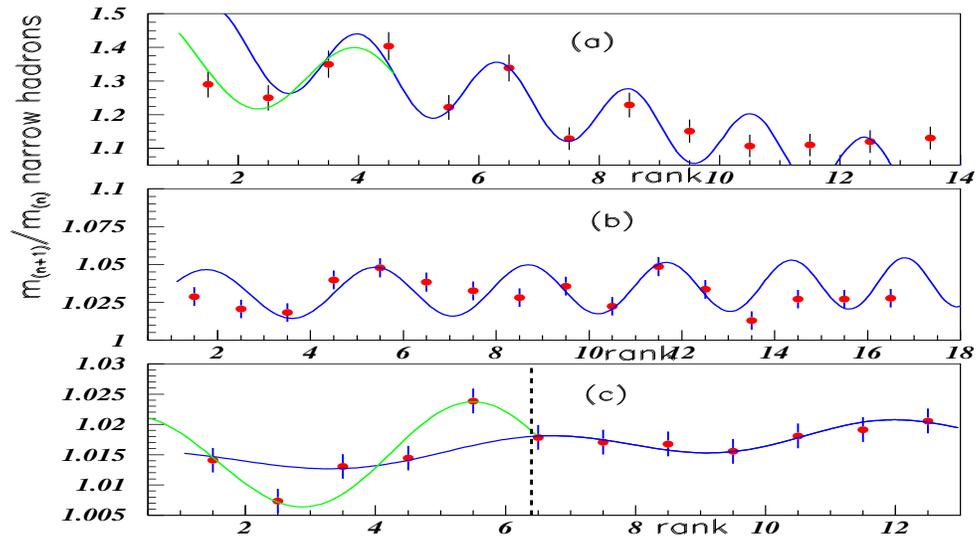}}
\caption{Log-log distributions of narrow exotic meson masses (insert (a)), narrow exotic baryon masses (insert (b)), and narrow exotic dibaryon masses (insert (c)) (see text).} 
\end{center}
\end{figure}

Fig.~9 shows the successive mass ratios of the same data (and references) as those used in fig.~8. The fits corresponding to the first ranks are improved by the introduction of a second set of parameters leading to green curves. Arbitrary relative errors are introduced in this figure, namely 1.5/100 for insert (a) data, 0.3/100 for insert (b) data, and 0.1/100
for insert (c) data. The data for mesons and dibaryons are better fitted  than the baryonic data.
\begin{figure}[ht]
\begin{center}
\scalebox{1.3}[0.8]{
\includegraphics[bb=40 170 526 520,clip,scale=0.6]{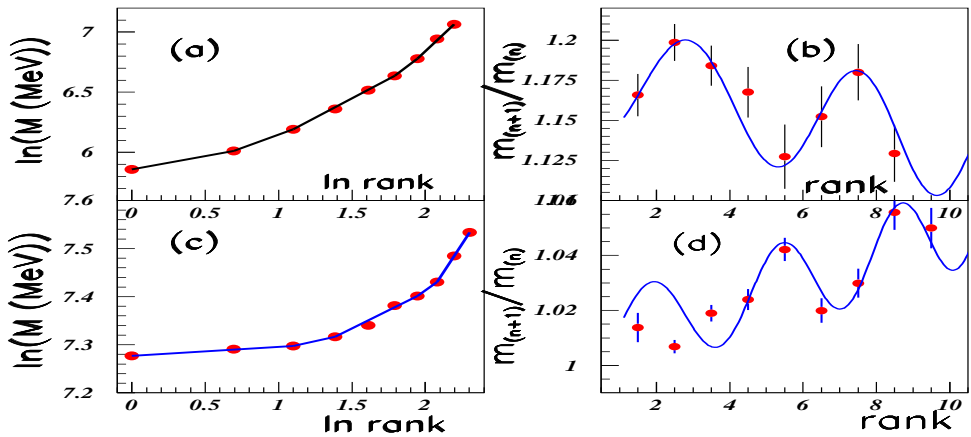}}
\caption{Log-log distributions of narrow exotic meson and baryon masses (see text).} 
\end{center}
\end{figure}

Fig.~10 shows in inserts (a) and (b) the log-log and successive mass ratios of the M$_{\pi^{+}\pi^{-}}$ invariant mass, from the np$\to$np$\pi^{+}\pi^{-}$   reaction measured at Dubna \cite{troyan1}. 
Inserts (c) and (d) show the log-log and successive mass ratios of the M$_{nK^{+}}$
invariant mass, from the np$\to$npK$^{+}K^{-}$ reaction measured at Dubna
 \cite{troyan2}. 

 In both cases, the log-log distributions exhibit straight line segments, and the successive mass ratios are well fitted.
 \begin{figure}[ht]
\begin{center}
\scalebox{1.3}[0.9]{
\includegraphics[bb=14 133 521 545,clip,scale=0.6]{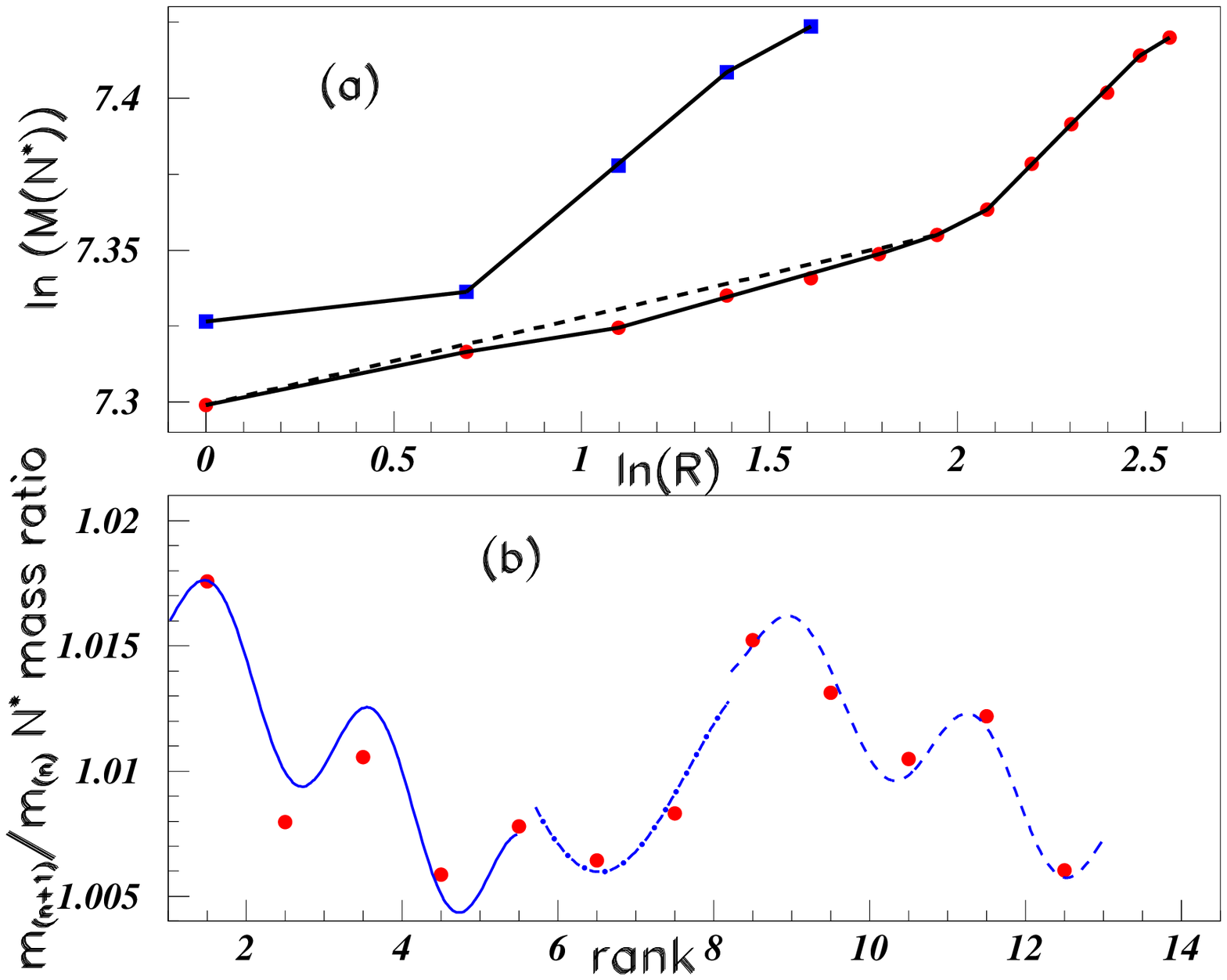}}
\caption{Log-log distributions of narrow $N^{*}$ structures in the mass range 1475$\le$ M $\le$ 1675~MeV (see text).} 
\end{center}
\end{figure}

An other example of narrow exotic baryonic structures concerns the careful study of the $N^{*}$ mass in the range 1475$\le N^{*} \le$1675~MeV. The same two reactions, already called: pp$\to$p$\pi^{+}$X and pp$\to$ppX were studied and the two final state channels: $p\pi^{0}$ and $p\eta$ were selected by software cuts applied to the missing mass of the first reaction \cite{bor6}. The conclusion of the work was that the broad PDG baryonic resonances in this mass range 
N(1520)D$_{13}$, N(1535)S$_{11}$, $\Delta$(1600)P$_{33}$, N(1650)S$_{11}$, and  N(1675)D$_{15}$ are collective states built from several narrow and weakly excited resonances,  having each a (much) smaller width than the one reported by PDG. Fig.~11(a) shows the log-log 
distribution of the broad (PDG) resonances in this mass range, by full blue squares and the log-log distribution of the 14 narrow resonances masses shown by full red circles. Notice again the possibility, shown by dashed line, to reduce the number of segments .

The straight lines include three masses in the case of broad PDG resonances and  (two times) five masses in the case of narrow masses.

Fig.~11(b) shows the values of the successive mass ratios of the narrow masses, fitted with equation (2.4) and three sets of parameters. 

\subsection{Discussion of the data extracted from the hadronic studies using fractals}
The variation of the parameters extracted from the fits of the successive mass ratios distributions versus the hadronic masses of all meson and baryon families, are shown in fig.~12.
\vspace*{-0.2cm}
\begin{figure}[ht]
\begin{center}
\scalebox{1.15}[1.1]{
\includegraphics[bb=14 186 520 520,clip,scale=0.7]{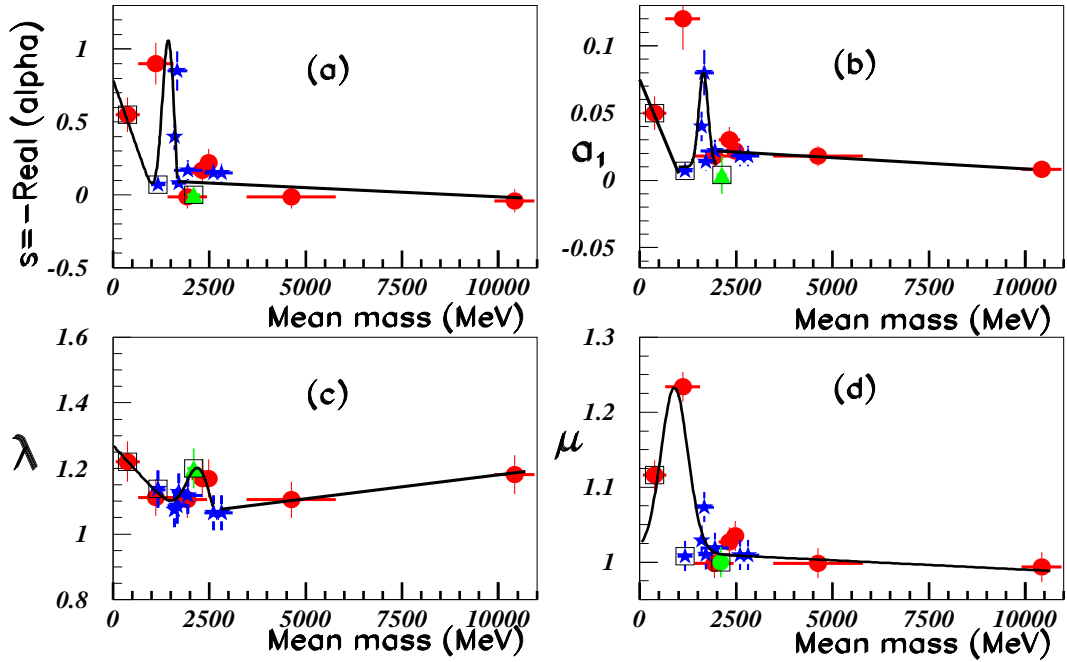}}
\caption{Variation of the three main fitted parameters versus the hadronic masses of all meson and baryon families. The meson data, the baryon data, and the dibaryon data are represented by red full points,  by blue full stars, and  by 
the green full triangles. The parameters encircled by black squares come from exotic narrow structure fits. The four inserts (a), (b), (c), and (d) correspond successively to -Re($\alpha$), oscillation amplitudes, a value related to Im($\alpha$), and a combination of $\lambda$ and "s", namely $\mu$ = $\lambda^{s}$.}  
\end{center}
\end{figure}
\vspace*{0.3cm}

The meson data are represented by red full points, the baryon data by blue full stars, and the dibaryon data by 
the green full triangles. The parameters encircled by black squares come from exotic narrow structure fits. The horizontal lines show the mass range of the given family and species. The marks are drawn in the middle of these horizontal lines.

 Fig.12(a) shows the variaton of 
"s" = -Re($\alpha$) which determines the general slope. Fig.~12(b) shows  the variation of "$a_{1}$" which determines the amplitudes of the oscillations. Fig.~12(c) shows the variation of $\lambda$ related to Im$(\alpha)$ by Im$(\alpha)$ = 2$\pi$n/ln$(\lambda)$. Fig.~12(d) shows  $\mu$ = $\lambda^{s}$.

We observe that each coefficient has a single variation, the same for all families and all species.

We observe also that Im($\alpha$) is $\gg$ that Re($\alpha$) in agreement with the DIS. Indeed the imaginary coefficient 2$\pi$/ln$\lambda$ = 66 for $\lambda$ = 1.1.
\subsection{Comparison of the excited hadronic state masses}
The distributions of the successive mass ratios  corresponding to various baryon and meson family distributions exhibit similar shapes. That suggests to compare the excited state masses of the various families. 

Fig.~13 shows the first PDG masses of different meson families, up to M = 3500~MeV, after a constant vertical mass translation of each 
family \cite{bor2}. Each column corresponds to a given family, arranged in increasing yrast masses (fundamental masses). The amount of the mass translations is indicated at the top of the figure as well as the family quark content. Therefore all yrast masses are put at the same value. The global scale corresponds to the charmed meson family (no mass translation).
\begin{figure}[ht]
\vspace*{-0.2cm}
\begin{center}
\scalebox{1.3}[1.3]{
\includegraphics[bb=4 276 520 520,clip,scale=0.6]{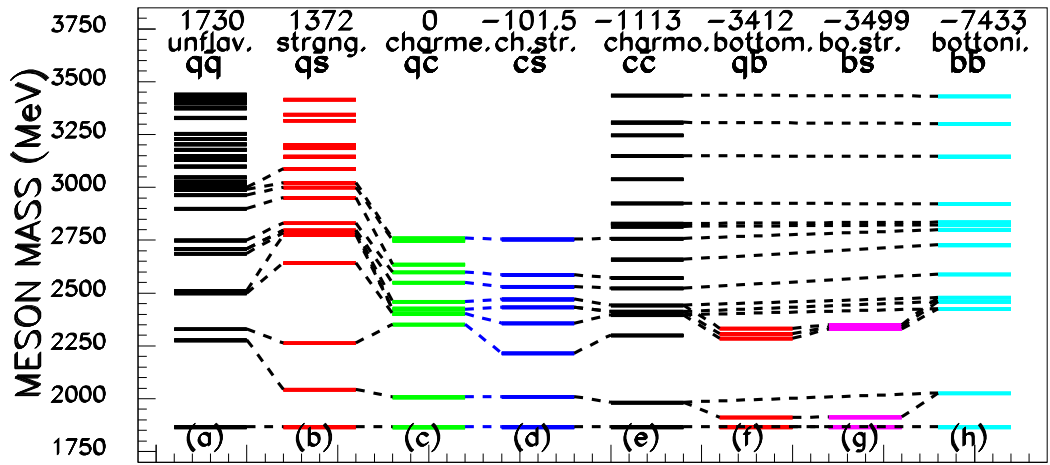}}
\caption{Comparison of meson excited state masses after a global mass translation to get the same value for the yrast masses (see text). q stands for u or d quarks.}  
\end{center}
\end{figure}

We observe, after translation, close continuous mass variations. For example  we observe very stable mass excitations between all three families containing a (two) charmed quark(s).
We also observe similar masses (after translation) between charmonium and bottonium mesons. These masses are joined by dashed lines. 
It is then possible to use this observation to tentatively predict some still unobserved masses shown by dashed lines \cite{bor2}. 
For example the following bottonium meson masses: 
M $\approx$ 9800~MeV, 10070~MeV, 10458~MeV, and 10660~MeV. And also a strange-charmed meson at M $\approx$ 2740~MeV.

Fig.~14 shows a similar discussion on the PDG baryonic families up to M = 2940~MeV \cite{bor2}.
\begin{figure}[ht]
\vspace*{-0.2cm}
\begin{center}
\scalebox{1.3}[1.3]{
\includegraphics[bb=4 276 520 520,clip,scale=0.6]{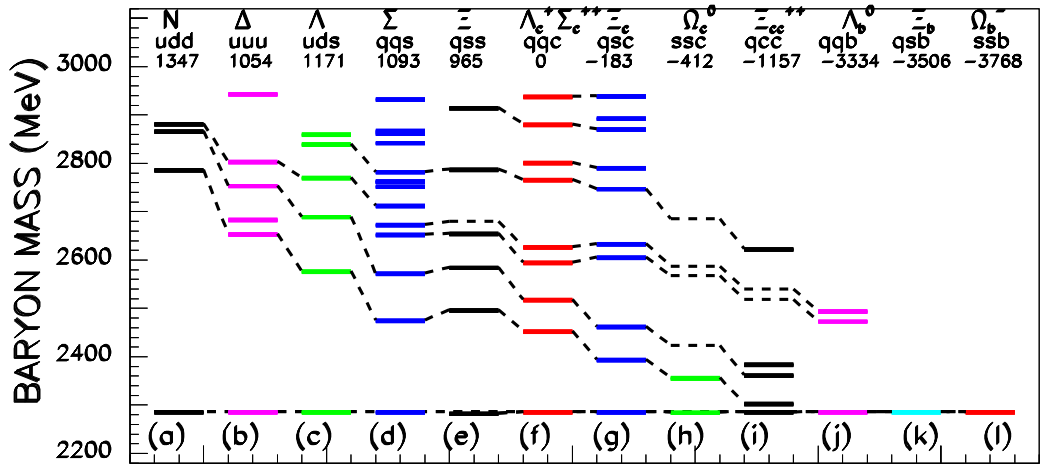}}
\caption{Comparison of baryon excited state masses after a global mass translation to get the same value for the yrast masses (see text).}  
\end{center}
\end{figure}

The comparable translation as the one performd on meson mass families, is done on baryon families. The global scale is adjusted to the charmed $\Lambda_{C}^{+}$ mass. We observe a regular mass decrease of the second and third masses of nearly all families, and also, although less regular, a mass decrease of the fourth and fifth masses. The shift of the excited masses of all families contract progressively when fundamental masses increase. Using this regularirty, several masses, still not observed, are tentatively predicted. They are shown in fig.~14 by dashed lines. These masses are:\\
\hspace*{0.2cm}M $\approx$ 1715~MeV for $\Xi$ baryons,\\
\hspace*{0.2cm}M $\approx$ 2850~MeV, 2980~MeV, 3000~MeV, and 3095~MeV for $\Omega^{0}_{C}$ baryons,\\
\hspace*{0.2cm}M $\approx$ 3750~MeV and 3775~MeV for $\Xi^{++}_{CC}$ baryons.

 When the baryon masses decrease more or less regularly after the described translations, the meson masses in the same conditions, seem to remain constant, after the strange ($q{\bar s}$) mesons.
\subsection{Study of the hadronic mass ratios}
The regular variation of the hadronic masses, discussed in the previous subsection, deserves more attention. 
In order to go more deeply inside these observations, we look at mass ratios between different hadronic families at the same rank \cite{bor5}. Fig.~15 shows the ratio between different baryonic families, the six inserts being explicited in Table~1. So the first (second ....) data in insert (a) shows the mass ratio between the first (second ....) $\Lambda_C, \Sigma_C$ mass over the first (second ....) $\Delta$ mass. We observe flat ratios in fig.~15, except the ratios between the yrast masses. That means constant interaction of QCD coupling constants and charges.
\vspace*{0.2cm}
\begin{figure}[ht]
\begin{center}
\scalebox{1.3}[0.85]{
\includegraphics[bb=18 65 525 545,clip,scale=0.6]{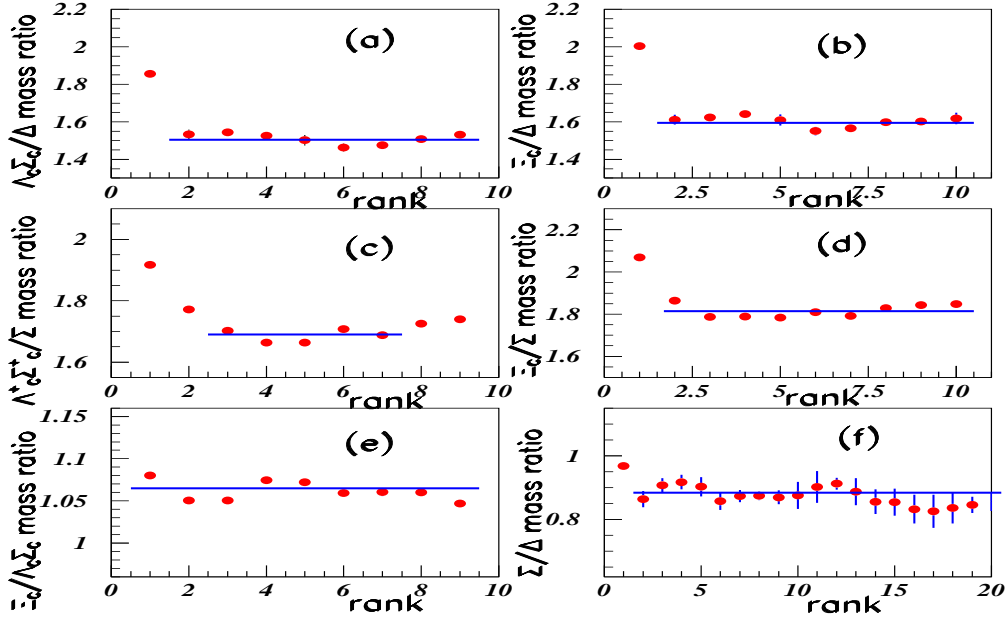}}
\caption{Mass ratios between different baryonic families (see Table 1).}  
\end{center}
\end{figure}
\vspace*{0.2cm}
\begin{table}[ht]
\begin{center}
\begin{tabular}[t]{c c c c c c}
\hline
{\bf (a)}&{\bf (b)}&{\bf (c)}&{\bf (d)}&{\bf (e)}&{\bf (f)}\\
\hline
{\bf ${\Lambda_{C}\Sigma_{C}/\Delta}$}&{\bf ${\Xi_{C}/\Delta}$}&{\bf ${\Lambda_{C}\Sigma_{C}/\Sigma}$}&
{\bf ${\Xi_{C}/\Sigma}$}&{\bf ${\Xi_{C}/\Lambda_{C}\Sigma_{C}}$}&{\bf ${\Sigma/\Delta}$}\\
${\bf udc,qqc/qqq}$&${\bf qsc/qqq}$&${\bf udc,qqc/qqs}$&${\bf qsc/qqs}$&${\bf qsc/udc,qqc}$&${\bf qqs/qqq}$\\
\hline
\end{tabular}
\caption{Table explaining the contents of the six inserts of fig.~15.}
\end{center}
\end{table}
The assumption is that this flat ratio property can be generalized between all baryonic families. This assumption is strengthened by the careful examination of the cases where the flat ratio is not observed.
When comparing the baryonic mass ratios where $N^{*}$ or $\Xi$ take place, we do no more have constant ratios in the total range available. Fig.~16 and Table~2 show the ratio of the excited masses of some baryonic families over the $N^{*}$ masses in insert (a), and the ratio of the $\Xi$ masses over the masses of some baryonic families in insert (b).
\begin{figure}[ht]
\begin{center}
\scalebox{1.3}[0.8]{
\includegraphics[bb=35 234 540 546,clip,scale=0.6]{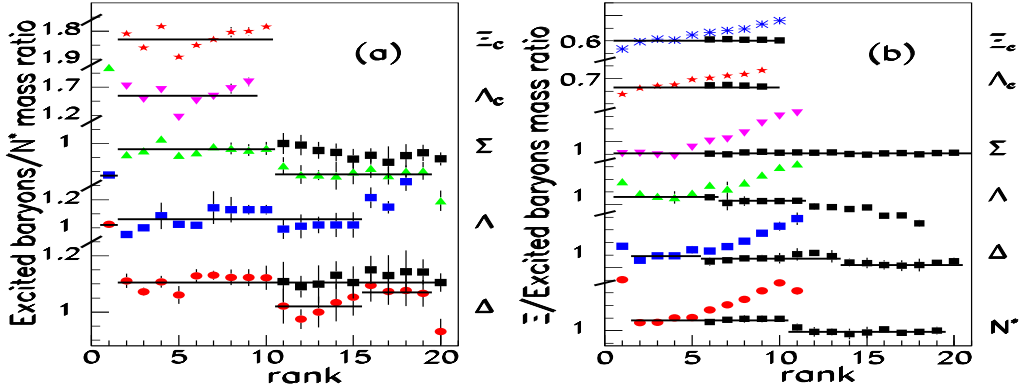}} 
\caption{Mass ratios between different baryonic families (see Table 2.)}  
\end{center}
\end{figure}
\begin{table}[h]
\begin{center}
\begin{tabular}[t]{c c c c c c}
\hline
{\bf (a)}&{\bf (b)}&{\bf (c)}&{\bf (d)}&{\bf (e)}&{\bf (f)}\\
\hline
{\bf ${\Xi_{C}/N^{*}}$}&{\bf ${\Lambda_{C}\Sigma_{C}/N^{*}}$}&{\bf ${\Sigma/N^{*}}$}&
{\bf ${\Lambda/N^{*}}$}&{\bf ${\Delta/N^{*}}$}&{\bf ${\Xi/\Xi_{C}}$}\\
${\bf qsc/udq}$&${\bf udc,qqc/udq}$&${\bf qqs/udq}$&${\bf uds/udq}$&${\bf qqq/udq}$&${\bf qss/qsc}$\\
\hline
{\bf (g)}&{\bf (h)}&{\bf (i)}&{\bf (j)}&{\bf (k)}\\
\hline
{\bf ${\Xi/\Lambda_{C}\Sigma_{C}}$}&{\bf ${\Xi/\Sigma}$}&{\bf ${\Xi/\Lambda}$}&{\bf ${\Xi/\Delta}$}&{\bf ${\Xi/N^{*}}$}\\
${\bf qss/udc,qqc}$&${\bf qss/qqs}$&${\bf qss/uds}$&${\bf qss/qqq}$& ${\bf qss/udq}$\\
\hline
\end{tabular}
\caption{Table explaining the six inserts of fig.~16.}
\vspace*{0.2cm}
\end{center}
\end{table}
 We observe in fig.~16(a) a constant ratio up to rank 10, followed by a sudden decrease at the same rank fo the three ratios. Such behaviour can be explained, and corrected, with the assumption of missing $N^{*}$ masses in the range 1730$\le$M$\le$1890~MeV. Indeed, in such range no  $N^{*}$ is reported in PDG \cite{pdg}. 
\begin{figure}[ht]
\begin{center}
\scalebox{1.3}[0.8]{
\includegraphics[bb=34 87 520 547,clip,scale=0.6]{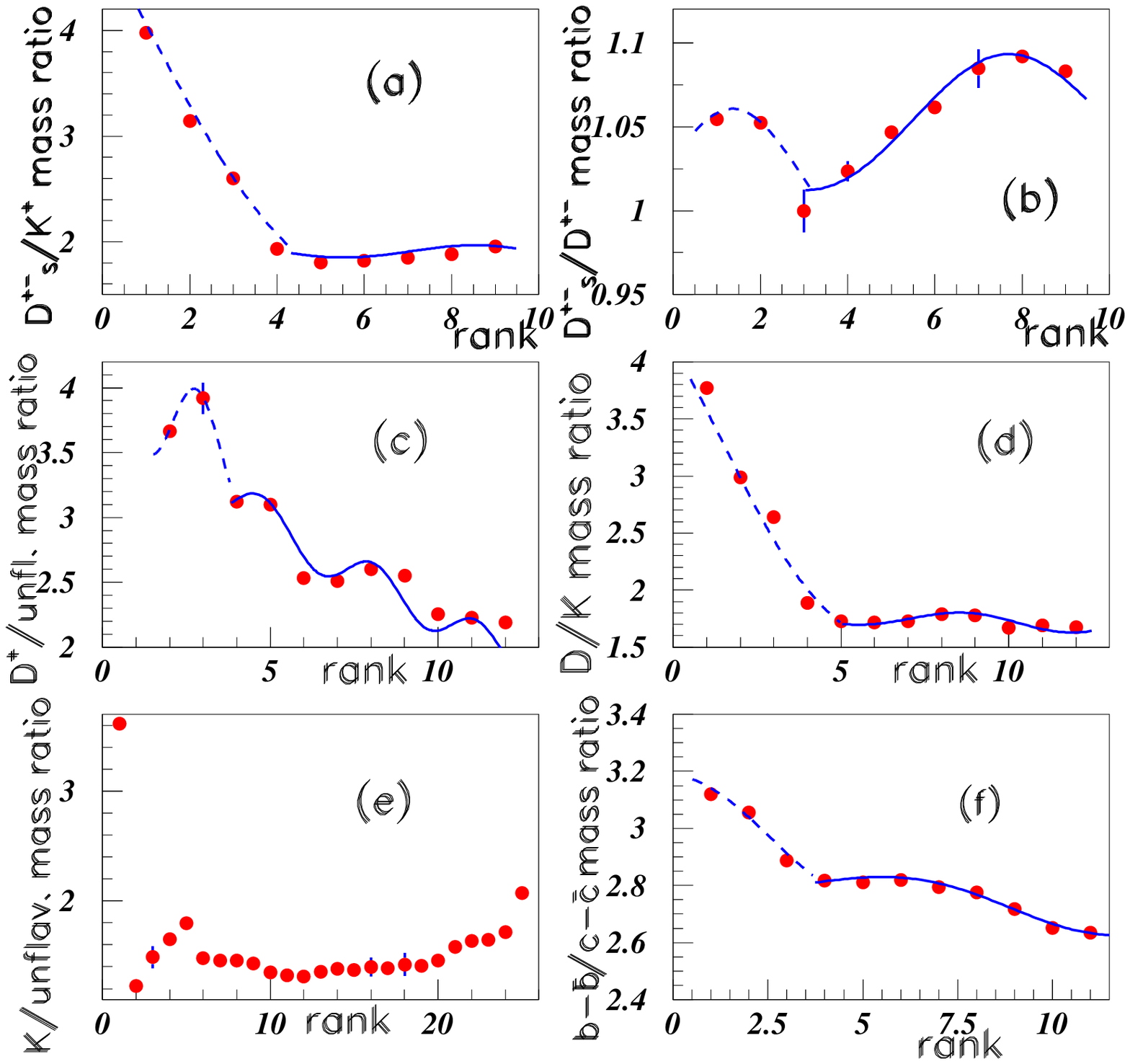}}
\caption{Mass ratios between different mesonic families explicited in table 3.}  
\end{center}
\end{figure}
The tentatively  introduction of three $N^{*}$ at M=1750, 1780, and 1820~MeV allows to find again the constant ratio, shown by black squares (correcting the green and red ratios after rank 10), between all shown families.
\begin{figure}[ht]
\begin{center}
\scalebox{1.3}[0.8]{
\includegraphics[bb=16 66 531 546,clip,scale=0.6]{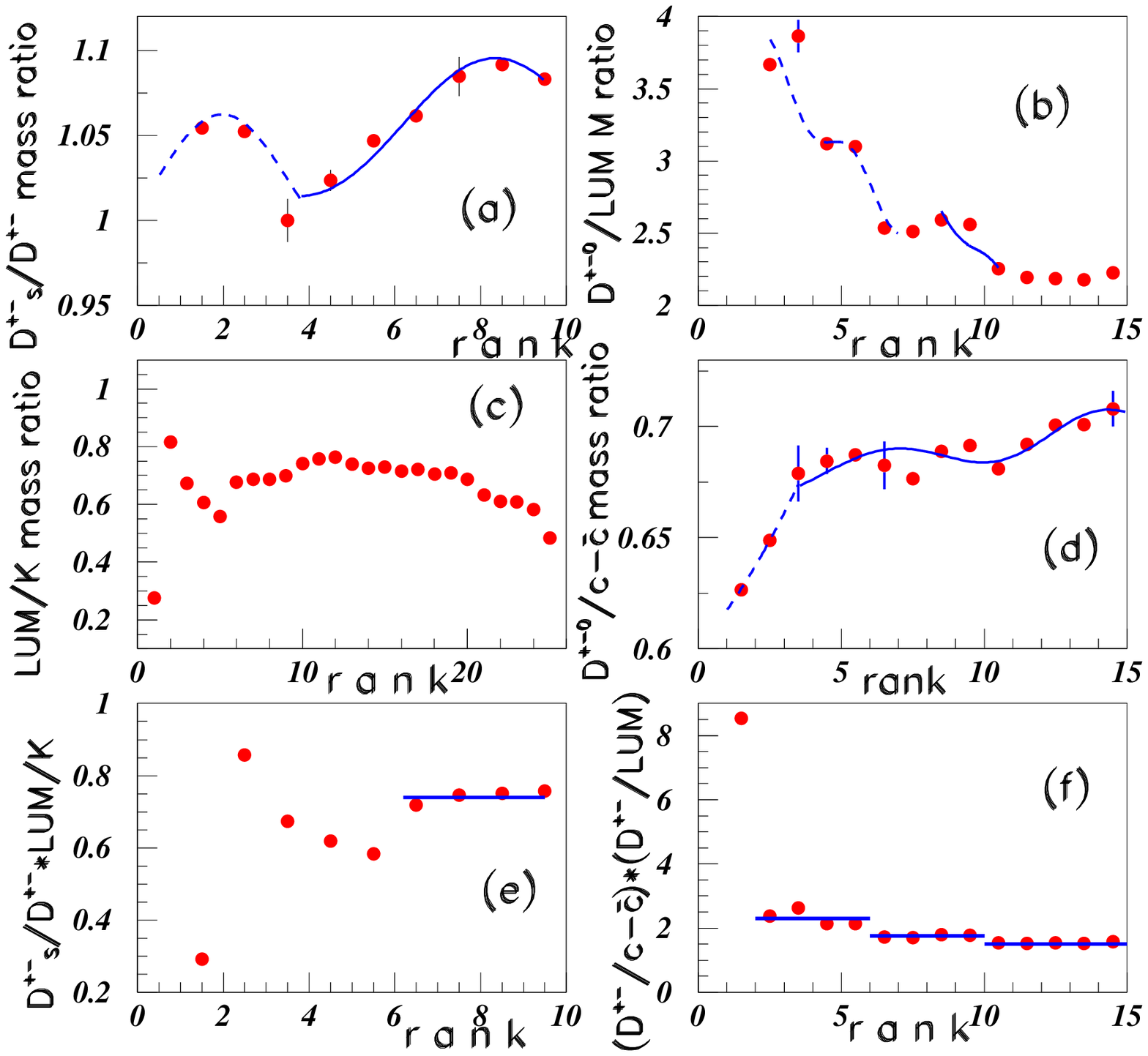}}
\caption{Double mass ratios between different mesonic families. Table 4
explicit the quark contents of the six inserts.}  
\end{center}
\end{figure}

The same discussion is done in fig.~16(b) and Table~3. Here the same arbitrary introduction of 14 $\Xi$ masses \cite{bor2} allows to get a constant
ratio in the whole range studied, for all families, shown by full black squares, when it was only observed for the first five ranks previously.

Similar study, performed between mesonic families, leads to a different result. Fig.~17 shows that the ratio of successive masses from different families, is not flat. Table~3 clarifies the contents of the six inserts.

 It is untimely to attribute the different behaviour between baryons and mesons to the introduction of more complex quark and/ or gluon configurations.
\begin{table}[h]
\begin{center}
\begin{tabular}[t]{c c c c c c}
\hline
{\bf (a)}&{\bf (b)}&{\bf (c)}&{\bf (d)}&{\bf (e)}&{\bf (f)}\\
\hline
{\bf ${D^{+-}_{S}/K^{+-}}$}&{\bf ${D^{+-}_{S}/D^{+-}}$}&{\bf ${D^{+-}/LUM}$}&
{\bf ${D^{+-}/K^{+-}}$}&{\bf ${K/LUM}$}&{\bf ${b{\bar b}/c{\bar c}}$}\\
${\bf c{\bar s}/q{\bar s}}$&${\bf c{\bar s}/c{\bar q}}$&${\bf c{\bar q}/q{\bar q}}$&${\bf c{\bar q}/q{\bar s}}$&${\bf q{\bar s}/q{\bar q}}$&${\bf b{\bar b}/c{\bar c}}$\\
\hline
\end{tabular}
\caption{Table explaining the contents of the six inserts of fig.~17.}
\end{center}
\end{table}
\begin{table}[h]
\begin{center}
\begin{tabular}[t]{c c c c c c}
\hline
{\bf (a)}&{\bf (b)}&{\bf (c)}&{\bf (d)}&{\bf (e)}&{\bf (f)}\\
\hline
{\bf ${D^{+-}_{S}/D^{+-}}$}&{\bf ${D^{+-}/LUM}$}&{\bf ${LUM/K^{+-}}$}&
{\bf ${D^{+-}/c{\bar c}}$}&{\bf ${D_{S}/D * LUM/K}$}&{\bf ${D/c{\bar c} * D/LUM}$}\\
${\bf c{\bar s}/q{\bar c}}$&${\bf c{\bar q}/q{\bar q}}$&${\bf q{\bar q}/q{\bar s}}$&${\bf c{\bar q}/c{\bar c}}$&${\bf c{\bar s}/q{\bar c}}$ * ${\bf q{\bar q}/q{\bar s}}$ = 1&${\bf q{\bar c}/q{\bar q}}$ * ${\bf c{\bar q}/c{\bar c}}$ = 1\\
\hline
\end{tabular}
\caption{Table explaining the contents of the six inserts of fig.~18.}
\end{center}
\end{table}

In order to try to understand such behaviour, fig.~18 and Table~4 show double ratios between mesonic mass spectra, performed in order to eliminate the quark masses. 
Here, insert (e) is obtained by the multiplication of insert (a) by insert (c). In quark world, we have $c{\bar s}/q{\bar c} * q{\bar q}/q{\bar s}$ = 1. In the same way, the right column corresponds to $q{\bar c}/q{\bar q} * c{\bar q}/c{\bar c}$ = 1 in fig.~18(f) obtained by multiplication of the content of fig.~18(b) by the content of fig.~18(d).  We observe three flat ratios in fig.~18(f), except the ratio between yrast masses. But such flat ratio is only observed in the last four ratios in insert (e).

In conclusion to this study of the hadron mass ratios, we remain with a double query:\\
\hspace*{0.9cm}(a) why do the ratios between different baryonic families remain flat ?,\\
\hspace*{0.9cm}(b) why are the behaviours between baryons and mesons so different ?\\
It remains possible that some "low meson masses"  have escaped from actual observation, but it is unlikely that this could occur in most meson families.
\section{Application to some nuclei yrast level masses}
Fig.~19 shows the log-log distributions of the masses of some light nuclei \cite{wapstra}.
\begin{figure}[ht]
\begin{center}
\scalebox{1.3}[0.8]{
\includegraphics[bb=4 330 525 546,clip,scale=0.6]{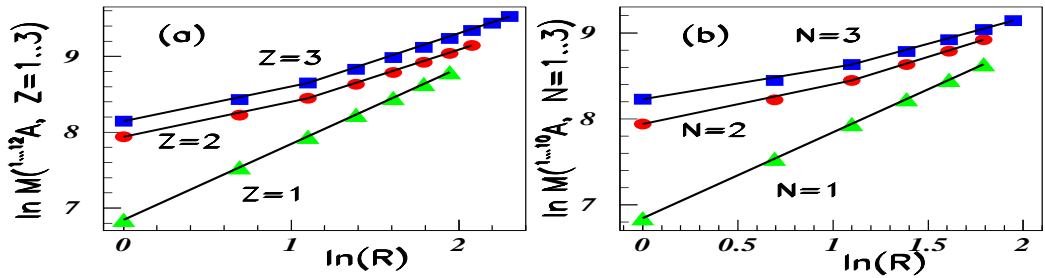}}
\caption[Light nuclei fractal properties]{Application to light nuclei Z (N) = 1, 2, and 3. Left side shows constant "Z" distributions for Z = 1 by full triangles (green on line), Z = 2 by full circles (red on line), and for Z = 3 by full squares (blue on line). 
The right side shows the comparable distributions for constant "N" nuclei. See Table 5. } 
\end{center}
\end{figure}
\begin{table}[h]
\begin{center}
\begin{tabular}[t]{c c c c c c c c}
\hline
      &           & last &exprim. &mass&next mass\\
Fig.&Marker&exper.& calcul&rel gap.&predicted\\
\hline
1&Z=1 &6568.98&6568.63&5.3 10$^{-5}$&$^{8}$H 7507.4\\
 &\hspace*{3.2mm}2 &9362.6&9244.4& 1.3\%&$^{11}$He 10179\\
 &\hspace*{2.8mm}3 &11226.3&11095.2&1.2\%&$^{13}$Li 12034\\
\hline
3 &N=1 &5629.9&5631.4&2.6 10$^{-4}$&$^{7}$C 6567.5\\
&\hspace*{3.2mm}2 &7483.8&7385.4&1.3\%&$^{9}$N 8312.3\\
&\hspace*{3.1mm}3 &9350.0&9204.4&1.6\%&$^{11}$O 10136\\
\hline
\end{tabular}
\caption[Attempt for mass extrapolations]{Experimentally observed masses and extrapolated next  masses with Z = 1, 2, and 3 nuclei, then N = 1, 2, and 3 nuclei corresponding to figure 19. Each line is concluded by the next predicted, but (still) not observed mass. All masses are in MeV. }
\end{center}
\end{table}

Nice alignments are observed, which precision is tested through the calculation of the masses of the last known nuclei masses, as shown in Table~5. The extrapolation is done using known masses at lower ranks, and ignoring the last known masses. These last ones are obtained with a mean precision close to 1.35$\%$ for N (Z) =2 or 3, and a much better precision for N (Z) = 1 (see Table 5). The next, therefore unknown masses, are therefore tentatively predicted.
\begin{figure}[ht]
\begin{center}
\scalebox{1.1}[0.7]{
\includegraphics[bb=1 142 527 552,clip,scale=0.7]{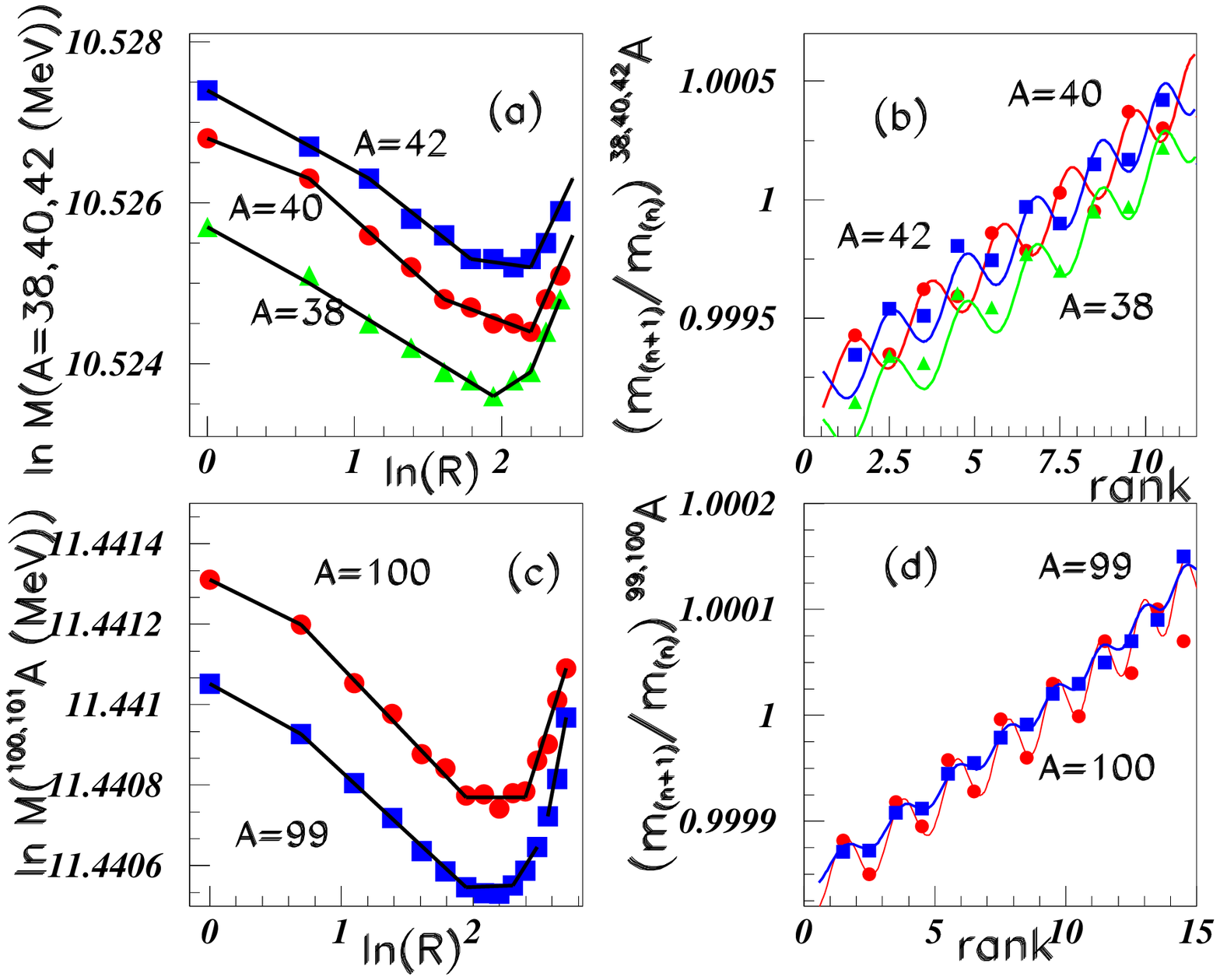}}
\caption{Application to A = 38, 40, 42, 99, and 100 nuclei (see text and table 6).} 
\end{center}
\end{figure}

Fig.~20 uses the variation of atomic masses for constant mass number and increasing number of protons (Z). Fig.~20(a) shows the log-log distribution of three nuclei masses for increasing Z values. Full red circles show data for A=40. Full blue squares show results for A = 42, the log values of these data are shifted by -0.048 in order to enter in the same figure. Full green triangles show results for A = 38; the log values of these last data are shifted by +0.0505 in order to enter in the same figure. 

Fig.~20(b) shows the $m_{n+1}/m_{n}$ mass ratios versus the rank, for nuclei close to A~=~40. Full red circles show data for A~=~40 and increasing Z values; full blue squares show results for A~=~42; full green triangles show results for A = 38. The experimental ratio values of these last data, and the calculated curve, are renormalized by 0.9998; otherwise they cannot be distinguished from A = 42 nuclei data. The fit gives rise to clearly observed nice oscillations, which describe the data in the whole experimental range.  All three curves are obtained using equation (2.4) and same parameters, except the phase $\Psi$. The A = 40 distribution is out of phase comparatively to the A = 38 and A = 42 nuclei since the first A = 40 nucleus is $^{15}P_{40}$ therefore with odd Z and N, when the first  A = 38 and A = 42 nuclei, are $^{14}Si_{28}$ and $^{16}S_{42}$, both having even Z and N. This observation shows the not usefulness to discuss the values of the phase "$\Psi$" of equation (2.4). These oscillations reproduce the pairing effect.   

Figs.~20(c) and 20(d) show the corresponding distributions for A = 99 and 100. Here Z increases, for both distributions, from 36 (rank 1) up to 50 (rank 15). We see that the many masses in the log-log plot of A~=~100 nuclei (full red circles) and in the log-log plot of A~=~99 nuclei (full blue squares), can be described by straight lines. The A~=~99 distribution is shifted by 0.00978 for clarity. For the A = 100 distribution, the atomic masses vary from $^{38}$Sr up to $^{48}$Cd.
These data should therefore follow the fractal properties.

Figs.~ 20(d) shows the $m_{n+1}/m_{n}$ mass ratio of A = 100 nuclei (full red circles)  versus the rank R and the  $m_{n+1}/m_{n}$ mass ratio of A = 99 nuclei (full blue squares). They both  exhibit many oscillations, describing the pairing effect. These oscillations are very well fitted by the equation~ (2.4) up to rank 15 (last rank). The oscillations are larger in case of even nuclei (A~=~100)  than in the case of odd nuclei (A~=~99). Indeed  for even A, the successive nuclei have "N" and "Z" alternatively both even and both odd, which is not the case for odd nuclei. The ratio between both a$_{1}$ factors, which describe the oscillation amplitude, is equal to 2.6. 
 
The linearity at the end of the distributions shown in figure~20(a), allows us to tentatively extrapolate the masses and predict the masses of still unobserved nuclei.
Table~6 gives the last experimental observed masses of the A = 38, 40, and 42 nuclei, compared to the  extrapolated masses. The extrapolated masses use the linearity of the previous masses in the log-log distribution. The relative errors between both lie between 1 10$^{-4}$ and  
2 10$^{-4}$. The last column shows the predicted mass of the next, still unobserved nuclei.
\begin{table}[h]
\begin{center}
\begin{tabular}[t]{c c c c c c}
\hline
      & last &exprim. &mass&next mass\\
Marker&exper.& calcul&rel gap.&predicted\\
\hline
A=38  {$\bigtriangleup$} & 35394.0 & 35397.6 & 1.0 10$^{-4}$
&  $^{38}$ V 35405\\
 \hspace*{4.2mm}40  {$\bigcirc$} & 37257.9 & 37249.7 &  2.1 10$^{-4}$ &                $^{40}$Cr 37268\\
 \hspace*{2.8mm}42 {$\Box$} & 39115.8 & 39108.7 & 1.8 10$^{-4}$  &     $^{42}$Mg 39128\\
\hline
\end{tabular}
\caption{Experimental last observed masses (in MeV) for $^{38, 40, 42}$A nuclei corresponding to figure~20, followed by the next predicted (extrapolated), but still not observed masses.}
\end{center}
\end{table}
\begin{figure}[ht]
\begin{center}
\hspace*{-3.mm}
\scalebox{1.2}[1.0]{
\includegraphics[bb=12 331 520 546,clip,scale=0.6]{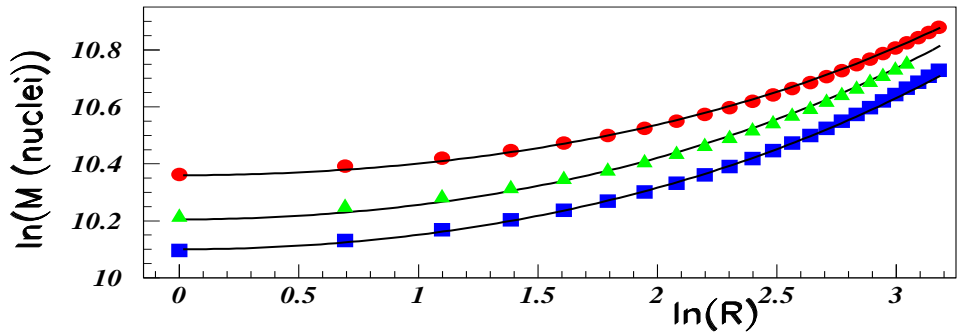}}
\caption{Log of some nuclei masses around $^{40}$Ca versus the log of the rank "R". The fig.  shows the distribution of calcium isotopes (full red circles), the distribution of sulfur isotopes (full blue squares), and the distribution of 
N~=~20 isotones (full green triangles), slightly shifted (see text).}
\end{center}
\end{figure}
\begin{figure}[ht]
\begin{center}
\hspace*{-3.mm}
\scalebox{1.2}[.75]{
\includegraphics[bb=4 140 522 545,clip,scale=0.7]{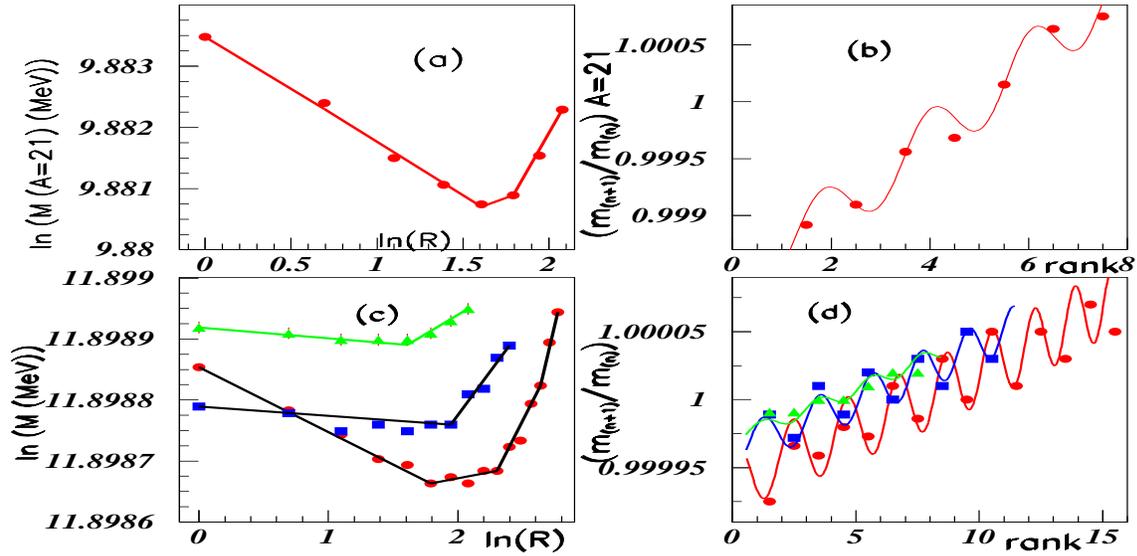}}
\caption{The four inserts (a), (b), (c), and (d) show respectively the log-log distribution of A = 21 nuclei versus increasing Z, the $m_{n+1}/m_{n}$ corresponding distribution, the log-log of the A = 158 nuclei (full circles red) and the log-log distribution of the A = 192 nuclei (full circles blue) and the A = 251 nuclei (full triangles green), and finally the corresponding successive mass ratios.}
\end{center}
\end{figure}

Figure 21 shows several log-log distributions for constant Z nuclei around $^{40}$Ca.  The distribution of calcium isotopes is shown by full red circles, the distribution of sulfur isotopes by full blue squares, and the distribution of N~=~20 isotones by full green triangles. These last data are shifted by +0.05 in order to clarify the fig.  

 There is no clear straight line segments here before rank 12; this fig. shows therefore that the linearized fractal model (LFM) considered up to now, is not  always observed. The data in fig.~21 are fitted by second order polynomials showing that they obey to parabolic fractal model (PFM) \cite{forriez}. The three parameters are now, successively for calcium (red data), sulfur (blue data), and N = 20 isotones (green data) the following. Order zero: 10.36, 10.205, and 10.1. Order 1: 0.04, 0.05, and 0.05. order two: 0.0011, 0.001, 0.001. 

Fig.~22 shows in the four inserts (a), (b), (c), and (d), respectively the log-log distribution of A = 21 nuclei versus increasing Z, the $m_{n+1}/m_{n}$ corresponding distribution, the log-log of the A = 158 nuclei (full circles red) and the log-log distribution of the A = 192 nuclei (full circles blue) and the A = 251 nuclei (full triangles green), and finally the corresponding successive mass ratios. The data exhibit oscillations which reduce for increasing mass. Here again, the experimental data are well fitted by equation (2.4). This pairing effect decreases again for increasing mass nuclei.
\begin{figure}[ht]
\begin{center}
\scalebox{1.3}[0.9]{
\includegraphics[bb=20 230 523 548,clip,scale=0.6]{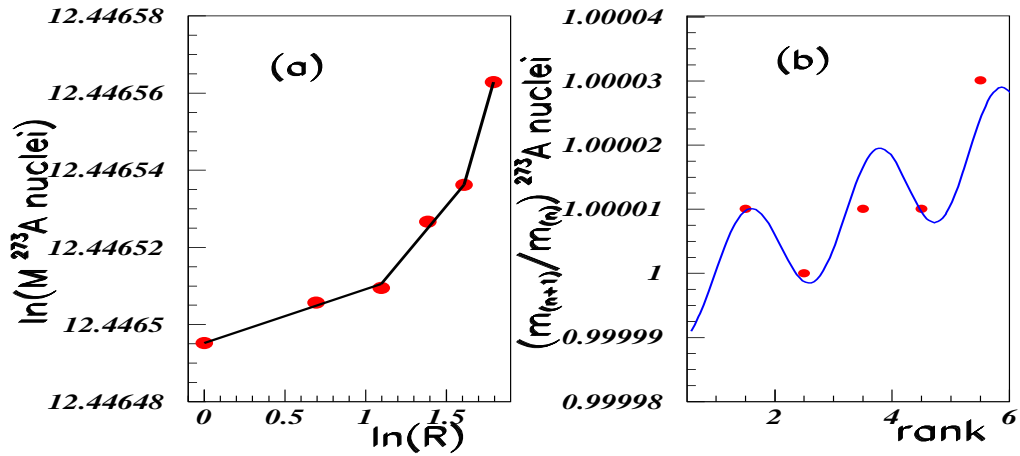}}
\caption{Log-log (insert (a)) and  $m_{n+1}/m_{n}$ distributions (insert (b)) of A = 273 nuclei versus increasing proton number "Z" from Z = 106 (rank 1) up to Z = 111 (rank 6).} 
\end{center}
\end{figure}

Fig.~23 shows the log-log (insert (a) and the $m_{n+1}/m_{n}$ distribution (insert (b)) of A = 273 nuclei versus increasing "Z" from Z = 106 (rank 1) up to Z = 111 (rank 6). 
\subsection{Study of the parameter's values}
\begin{figure}[ht]
\begin{center}
\scalebox{1.3}[0.9]{
\includegraphics[bb=58 237 528 546,clip,scale=0.6]{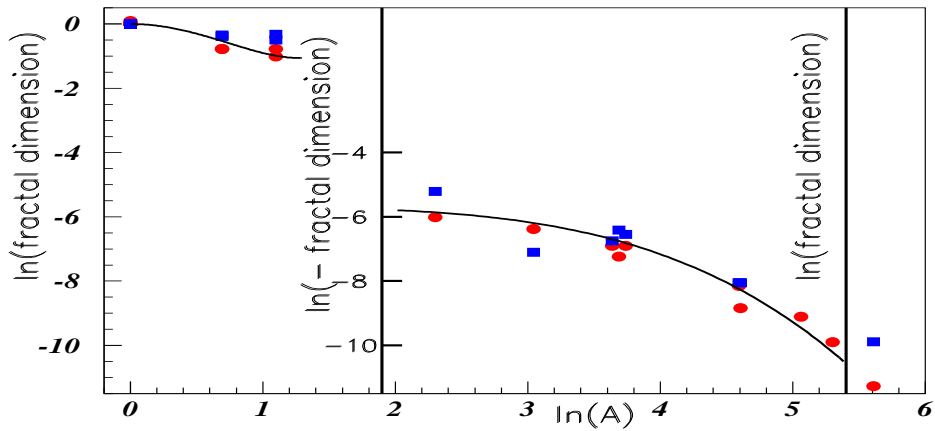}}
\caption{Slopes of the two first straight line segments versus the log of the atomic number (see text).} 
\end{center}
\end{figure}

We start with the dicussion on the fractal dimension values: the slopes of the straight line segments in the log-log distributions. As seen before, the number of such segments varies from one up to four, mainly equal or larger than two. Fig.~24 shows the slopes of the two first straight line segments versus the log of the atomic number. 
Red full circles correspond to the first straight line slopes, blue full squares correspond to the second straight line slopes.
Since these slopes decrease fast, and are quickly negative, they are shown in the fig.~24, either through  log(d),  or through the log(-d).  In spite of a dispersion, we observe a global regular decreasing variation for increasing masses.}

The fits performed on many successive mass ratio nuclei, determine many different values of the three main parameters. It is important to verify their continuous non random variation. 
\begin{figure}[ht]
\begin{center}
\scalebox{1.3}[0.9]{
\includegraphics[bb=49 137 526 551,clip,scale=0.6]{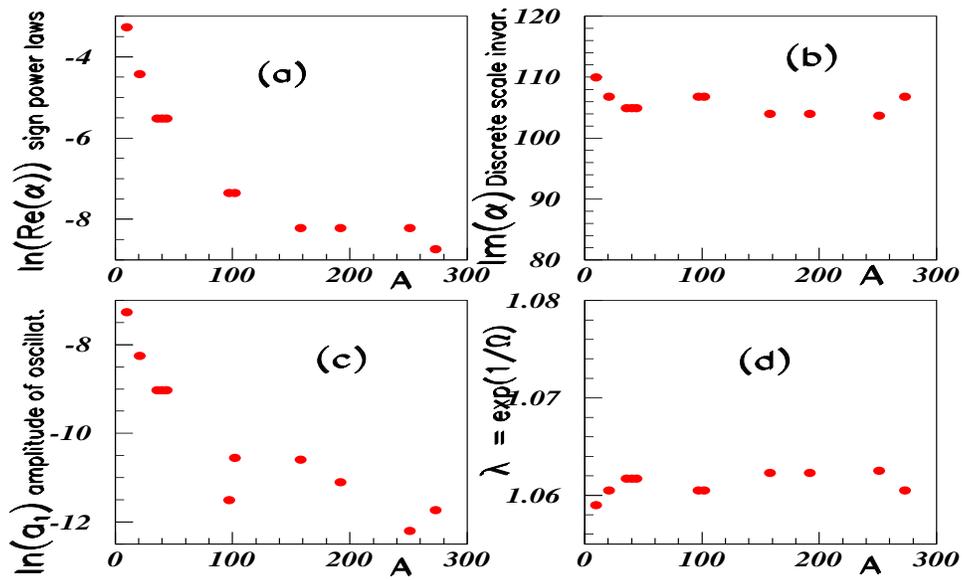}}
\caption{Parameters which fit the $m_{n+1}/m_{n}$ nuclei mass ratios (see text).} 
\end{center}
\end{figure}

This is done in fig.~25 which shows the variation of the parameters fitting the various $m_{n+1}/m_{n}$ nuclei mass ratios. Inserts (a), (b), (c), and (d) show respectively the variation of ln (Re$(\alpha)$), Im$(\alpha)$, 
ln ($a_{1}$), and $\lambda$. The last one (d) is simply the reflection of the second (b), since $\lambda = exp(1/\Omega)$. In the case of A = 251, the small number (5) of ratios, involve a less precise fit. The poor precision on $\Omega$ is also a consequence of a small oscillation amplitude since it is an odd (and large) mass number. All four inserts show continuous parameter variations. 
\subsection{Application to some lines or columns of the Mendeleev periodic table of elements}
\begin{figure}[ht]
\begin{center}
\scalebox{1.3}[1.0]{
\includegraphics[bb=36 240 542 548,clip,scale=0.6]{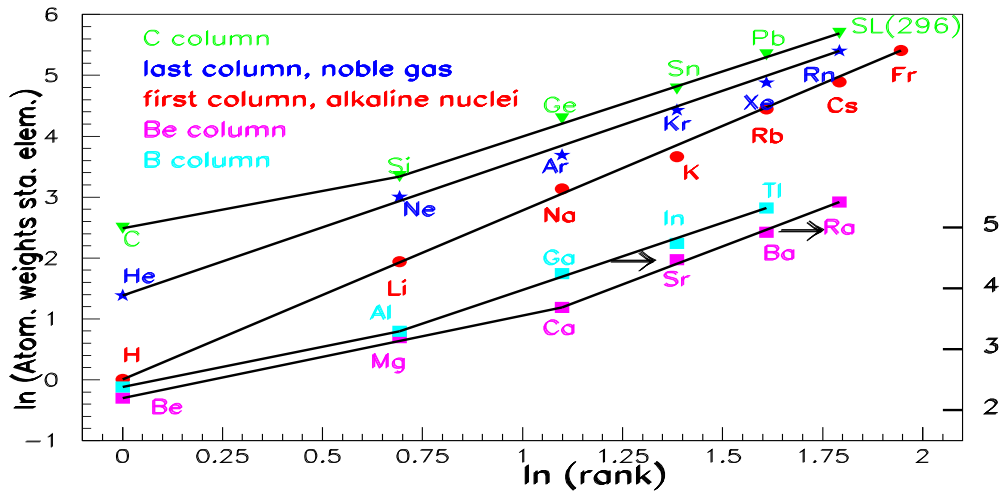}}
\caption{Log-log distributions of several columns of the Mendeleev periodic table of elements.} 
\end{center}
\end{figure}
\begin{figure}[ht]
\begin{center}
\scalebox{1.3}[0.93]{
\includegraphics[bb=20 235  539 551,clip,scale=0.6]{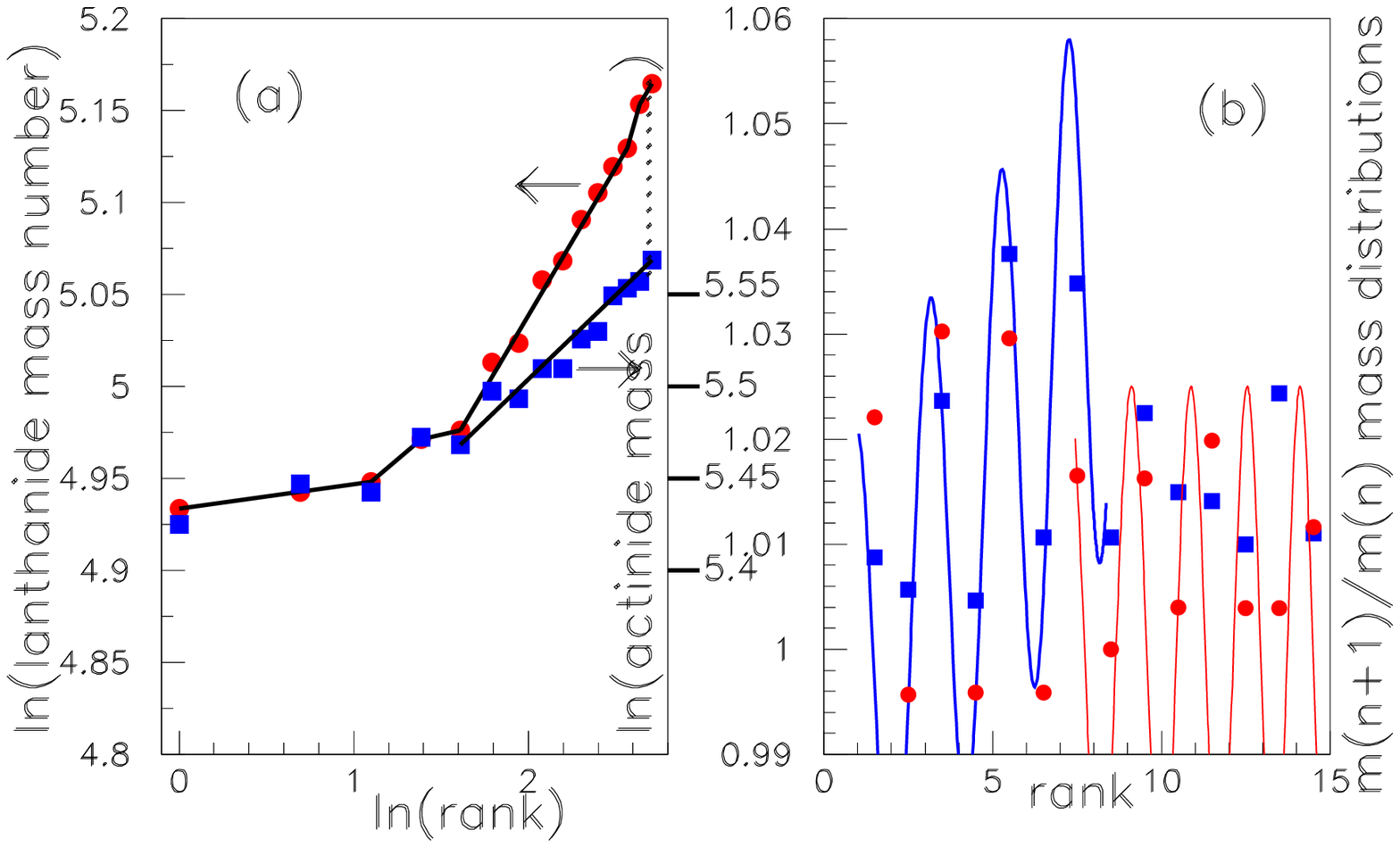}}
\caption{Log-log distributions of the Lanthanides and Actinides lines  of the Periodic table of elements (see text).} 
\end{center}
\end{figure} 
Fig.~26 shows the log log distribution of masses from some columns of the elements of the Mendeleev periodic table of elements. Only one straight line is observed to describe the first and last columns masses, when the other column distributions show two straight line segments. The left scale corresponds to the carbon (C), (He), and (H) columns, when the right scale corresponds to the (B) and (Be) columns. 

Fig.~27 shows the log-log distributions and $m_{n+1}/m_{n}$ distributions of the lanthanide serie (full red circles) and the actinide serie (full blue squares) \cite{pdg} with corresponding scale at the right part of the fig.~27. The log-log data for the first five ranks differ exactly by a translation which amount equals to 0.5. The $m_{n+1}/m_{n}$ distributions are fitted by the same two sets of parameters.
The second set of parameters, drawn in red with Re($\alpha$) = 0, differs from the first one (drawn in blue),  by the only one different parameter, namely Re($\alpha$) = -0.2.
\begin{figure}[ht]
\begin{center}
\scalebox{1.35}[0.86]{
\includegraphics[bb=1 91 521 555,clip,scale=0.6]{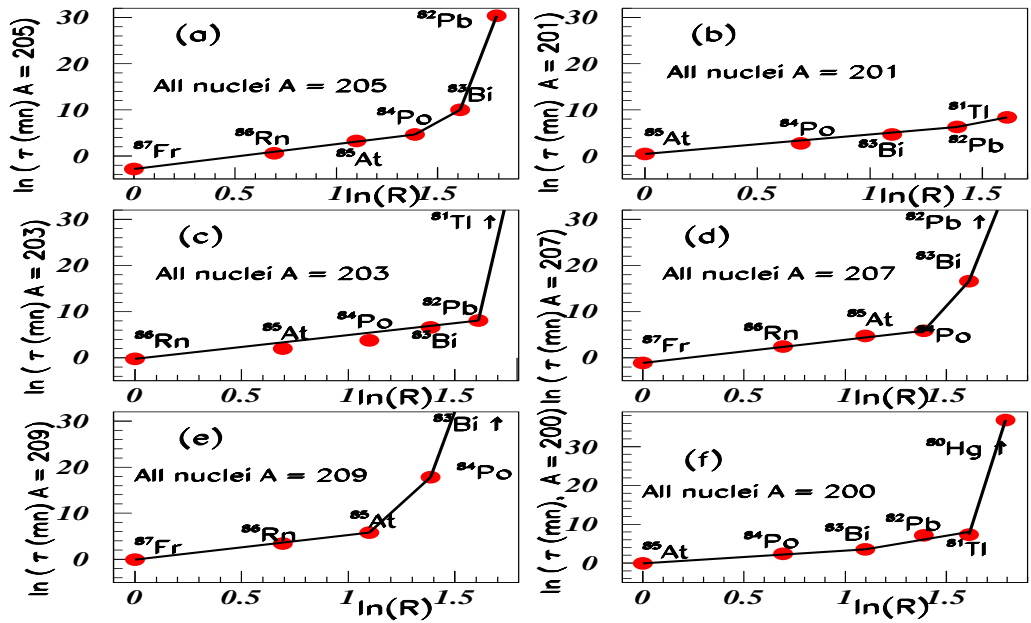}}
\caption{Log-log distributions of some mass series following [EC] or $\beta^{+}$ disintegrations.} 
\end{center}
\end{figure}
\subsection{Application to mass series following  $\beta^{+}$ or $\beta^{-}$ disintegrations.}
\begin{figure}[ht]
\begin{center}
\scalebox{1.3}[0.9]{
\includegraphics[bb=2 88 520 550,clip,scale=0.6]{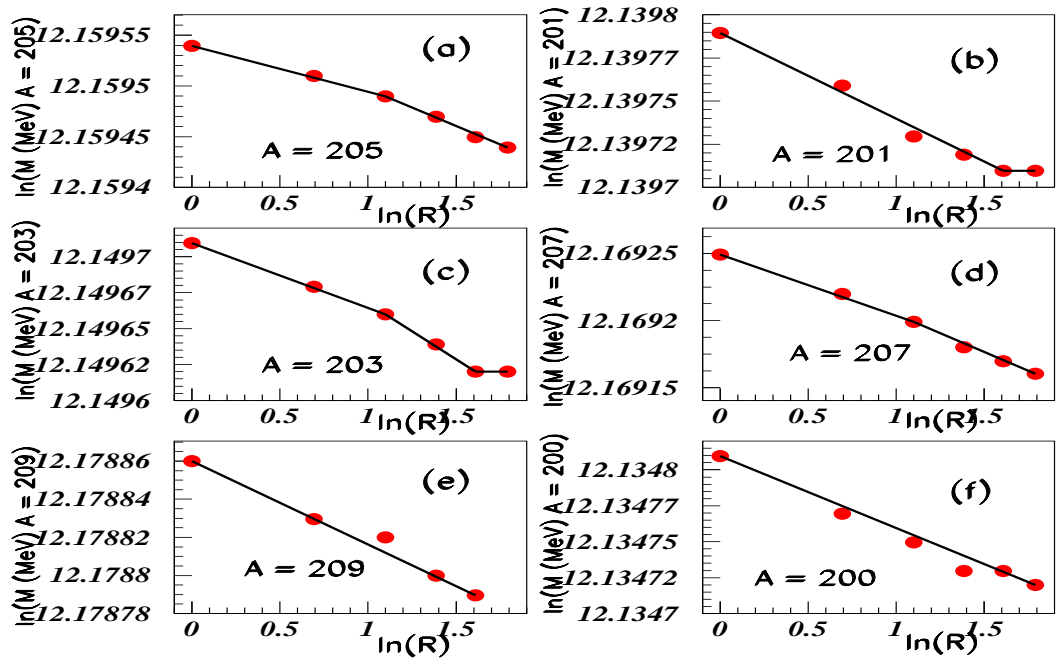}}
\caption{Log-log distributions of half-life series following [EC] or $\beta^{+}$ disintegrations.} 
\end{center}
\end{figure}
\begin{figure}[ht]
\begin{center}
\scalebox{1.3}[0.9]{
\includegraphics[bb=2 88 520 550,clip,scale=0.6]{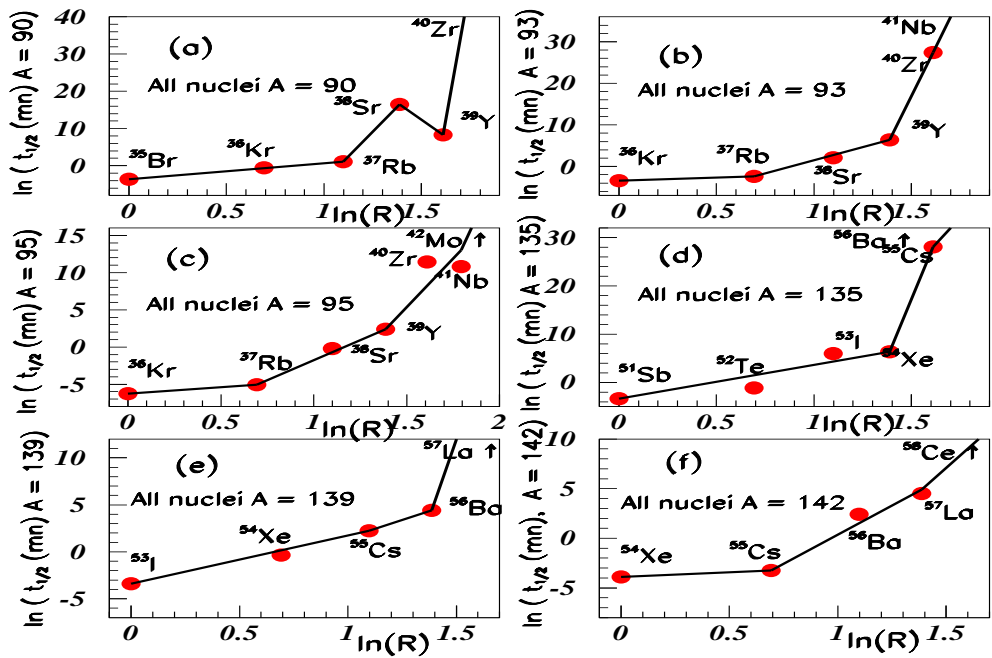}}
\caption{Log-log distributions of half-life series following  $\beta^{-}$ disintegrations (see text).} 
\end{center}
\end{figure}
\begin{figure}[ht]
\begin{center}
\hspace*{-3.mm}
\scalebox{1.3}[.72]{
\includegraphics[bb=19 130 539 544,clip,scale=0.6]{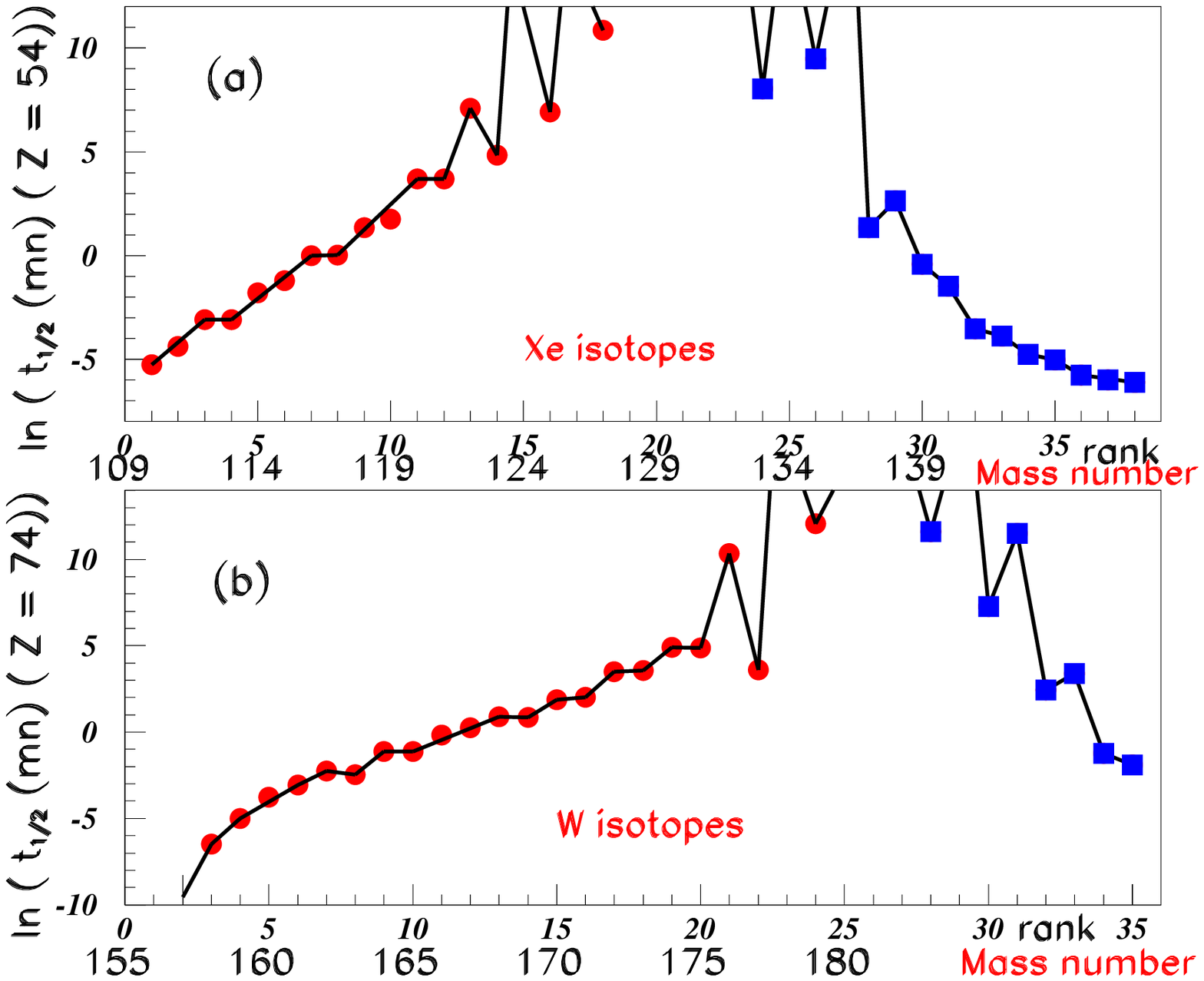}}
\caption{Application to half-lifes of Xe and W isotopes (see text).} 
\end{center}
\end{figure}

Fig.~28 shows the log-log distributions of the mass series following [EC] (electron capture) or $\beta^{+}$ disintegrations. The six inserts corresponds each to a constant mass number "A": 200$\le~A~\le$209, and decreasing proton number "Z".  For example, in insert (a), rank 1 corresponds to $^{87}$Fr and rank 6 corresponds to $^{82}$Pb. They all consists of at least 4 aligned data.

Fig.~29 shows some log-log distributions of the half-life series following [EC] or $\beta^{+}$ disintegrations. The six inserts corresponds to 200$\le~A~\le$209, the same nuclei than those shown in previous fig.~27.  We observe the same tendancy as before, although perhaps less pronounced.

Fig.~30 shows some log-log distributions of the half-life series following $\beta^{-}$ disintegrations. The six inserts corresponds to A = 90, 93, 95, 135, 139, and A = 142. Here the alignement is not as good, as it was observed previously.
\subsection{An example of log-periodic law: Half lifes of Xe and W isotopes}

Fig.~31 shows the half-life variations of the Xe and W isotopes and illustrates a log-periodic system, here applied to times. Half-lifes increase from a mass number A to A+1, by $\beta^{+}$ emission up to a critical mass number "A$_{C}$" (stable nucleus) \cite{nottale1} \cite{nottale2}. This evolution is represented by full red circles. Here there are several stable nuclei, (not shown in the figure), therefore several critical mass numbers.  After the "critical range", the half lifes decrease when the $\beta^{-}$ emission allows to go from A up to A+1 nuclei. Such evolution is characterized by an autosimilarity parameter "g"
defined by 
\begin{equation}
g = (A_{n} - A_{C})/(A_{n+1} - A_{C}). 
\end{equation}
The poor definition of "$A_{C}$, join to the systematic oscillation every successive data, involves a very imprecise value for "g" (from 1.15 up to 1.5).
\section{Application to nuclei excited level masses}
A  number of data, larger than previously for yrast masses, exist for the excited level energies of many nuclei. The excited level masses studied are introduced with increasing masses, starting from the fundamental one. The energy levels for different nuclei, are taken from several 
 F.~ Ajzenberg-Selove and T.~Lauristen papers.
\subsection{Application to excited level energies for nuclei masses A=11, 12, 13, and 14.}
\begin{figure}[ht]
\begin{center}
\hspace*{-3.mm}
\scalebox{1.3}[.9]{
\includegraphics[bb=6 230 530 550,clip,scale=0.6]{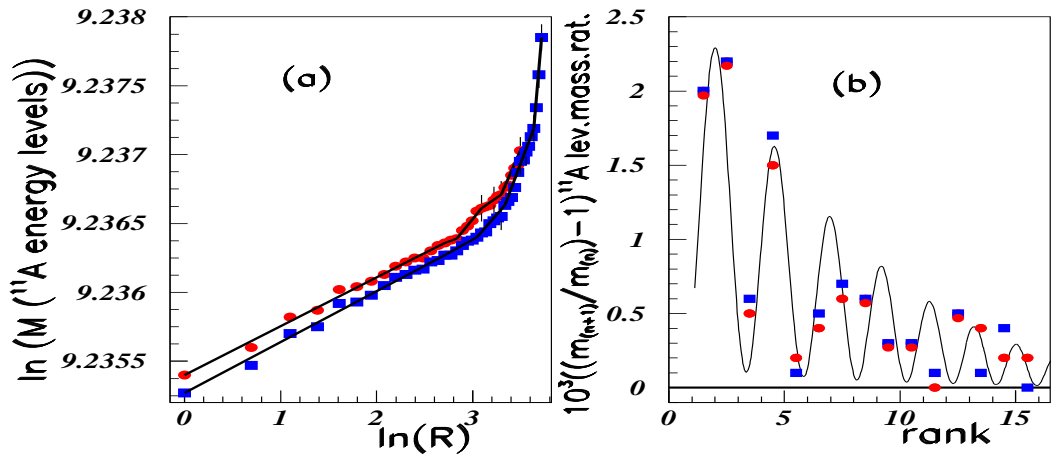}}
\caption{Log-log plots and successive mass ratios of excited level masses of $^{11}$C in full red circles and  $^{11}$B in full blue squares.} 
\end{center}
\end{figure}

We observe in fig.~32(a), nice and parallel alignements of log-log distributions for the first 16 levels of $^{11}$C (full red circles), and of $^{11}$B (full blue squares) \cite{as111213}. Fig.~32(b) shows the corresponding $m_{n+1}/m_{n}$ mass ratio distributions of $^{11}$C energy levels: (full red circles), and  $^{11}$B energy levels: (full blue squares). The same set of parameters, allows, using equation (2.4), to well reproduce these ratios for both nuclei at least up to rank 11. 
\begin{figure}[ht]
\hspace*{-3.mm}
\scalebox{1.3}[1.]{
\includegraphics[bb=4 235 520 547,clip,scale=0.6]{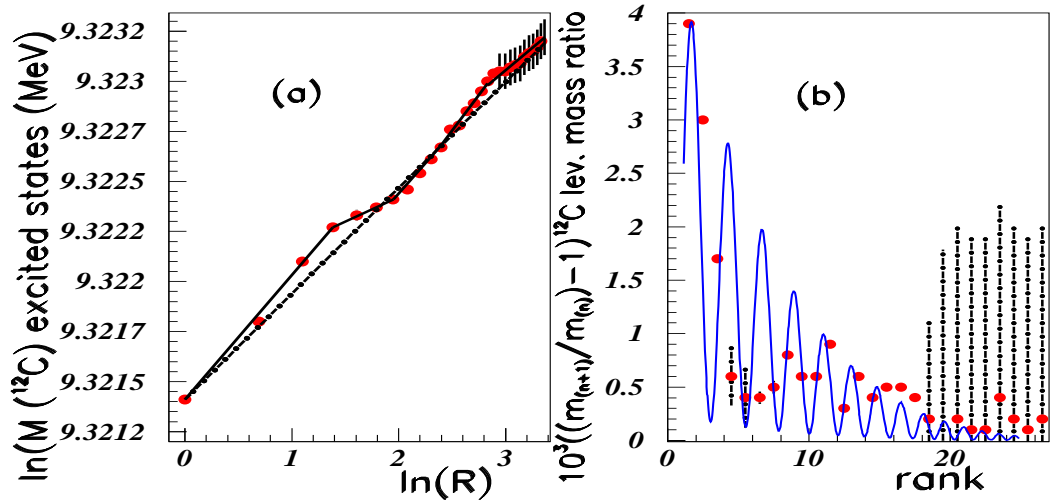}}
\caption{Log-log plots and successive mass ratios of excited level masses of $^{12}$C.} 
\end{figure}

Figure~33(a) shows the log-log plot of the $^{12}$C excited energy level masses \cite{as111213}.
We observe an alignement, with two slight shifts at the beginning (low "R"), and in the vicinicy of the 17$^{th}$ level (mass = 17.76~MeV).
Figure~33(b) shows the $m_{n+1}/m_{n}$ mass ratios between adjacent  excited energy $^{12}$C energy levels.
Since the relative mass  difference between adjacent levels is very small, a few MeV, versus more than 11 GeV for the total mass, we plot here 1000*((m$_{n+1}$/m$_{n}$)-1) instead m$_{n+1}$/m$_{n}$. When the mass precision is not given in the tables, we arbitrarily attribute 1 MeV for these uncertainties. The large error bars starting at 
R = 17.5 manifest these large error arbitrarily introduced. The distribution fits all the fifteen first excited level masses, although the 2$^{nd}$, 3$^{rd}$, and 4$^{th}$ calculated maximas have no experimental counterpart. The ratio of rank 6.5 is far from the maximum of the calculated distribution at the same rank.
\begin{figure}[ht]
\begin{center}
\hspace*{-3.mm}
\scalebox{1.3}[1.]{
\includegraphics[bb=6 230 530 550,clip,scale=0.6]{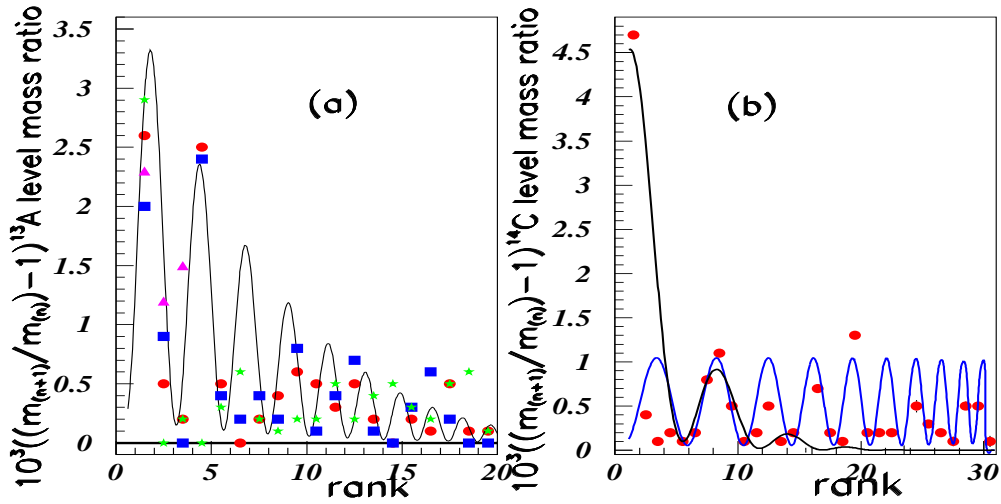}}
\caption{Log-log plot and successive mass ratio of energy level masses of $^{13}$A  and $^{14}$C nuclei (see text).} 
\end{center}
\end{figure}

Fig.~34(a) shows the $m_{n+1}/m_{n}$ mass ratios of excited levels for A~=~13 nuclei \cite{as111213}. Full red circles show the distribution for $^{13}$C nuclei, full blue squares show the distribution  for $^{13}$N nuclei,  full green stars show the distribution  for $^{13}$B nuclei,  and  full  purple triangles show the distribution  for $^{13}$O nuclei. The curve corresponds to a fit obtained using equation (2.4).
As before, we observe an unique and rather good fit (same parameters) for all four A = 13 nuclei, up to R~=~16, which spoils for larger R values. 
Fig.~34(b) shows the $m_{n+1}/m_{n}$ mass ratios of $^{14}$C levels \cite{as14}.
We observe here a case where the equation (2.4), with an unique set of parameters, does not allow to fit the experimental data. Two solutions are drawn, both describe well the "peak" around rank 8. The first solution (black on line), catches also the first large point, but not only forget the data after rank 11, but also does not describe the data for rank 2 and 3.  The second solution (blue on line) forgets completely the data points at ranks 2 and 3.

Several other excited excited level nuclei masses were studied through fractal presence \cite{bor3}. These nuclei are: $^{16}$O \cite{as16}, $^{23}$Na, 
$^{46}$Ti, $^{62}$Ni, $^{92}$Zr, and $^{134}$Ba. The corresponding results are omitted here. They can be summarized by the two already often observed properties: the log-log distributions exhibit a few straight line segments, and the fit for successive mass ratio distributions describe the data up to rank 6 to 8 with an unique set of parameters.
\subsection{Excitation level masses of some heavy nuclei.}
\begin{figure}[ht]
\begin{center}
\hspace*{-3.mm}
\scalebox{1.3}[1.]{
\includegraphics[bb=13 88 517 554,clip,scale=0.6]{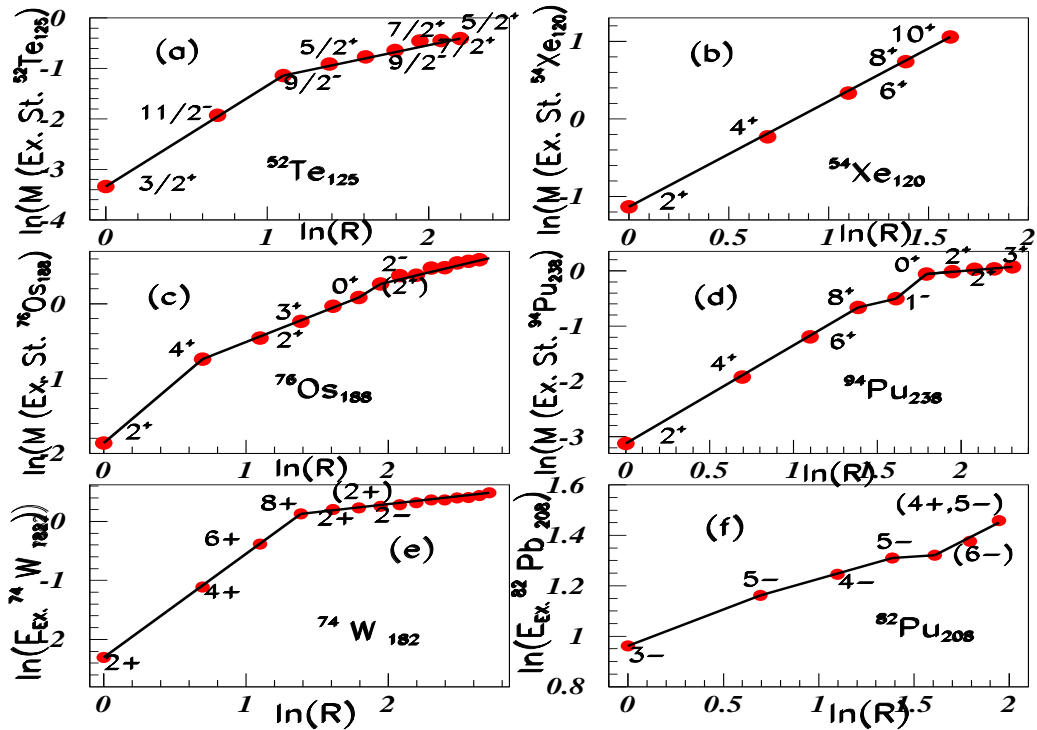}}
\caption{Log-log plots of excited level masses (in MeV) of some heavy nuclei (see text).} 
\end{center}
\end{figure}
Fig.~35 shows the log-log plots of six heavy nuclei excited level masses: $^{52}$Te$_{125}$, $^{54}$Xe$_{120}$, $^{76}$Os$_{188}$, $^{94}$Pu$_{236}$, $^{74}$W$_{182}$, and $^{82}$Pb$_{208}$ \cite{lederer}.
 Since the excitation energies are too small, compared to the masses, the log of the excitation energies are plotted and not the log of the total masses. It follows that the yrast masses are not present. 

 The spin values are reported on the figure, allowing to observe very nice alignements for the rotationnal spectra (inserts (b), (d), and (e)).
\subsection{Discussion on the parameter values extracted from fits on nuclei.} 
The parameter values (named a) extracted from excited  state nuclei masses (up to $^{62}$Ni), are different from those (named b) extracted from yrast mass variations. 
The critical exponent "s" is close to 5 (case (a)), but is very small and negatif ($\le - 10^{-2}$) in case (b). The amplitude of the log-periodic correction to continuous scaling parameter "a$_{1}$" is also much larger in case (a) close to 0.9, but very small ($\le 10^{-3}$) in case (b). 

The main parameter of these (DSI) is $\lambda$, directly related to Im($\alpha$). Fig.~36 shows the $\lambda$ values extracted from nuclei mass variations (yrast masses) drawn by purple triangles, and nuclei energy level mass variations drawn with sky blue stars.
 \begin{figure}[ht]
\begin{center}
\hspace*{-3.mm}
\scalebox{1.4}[1.5]{
\includegraphics[bb=28 336 517 518,clip,scale=0.5]{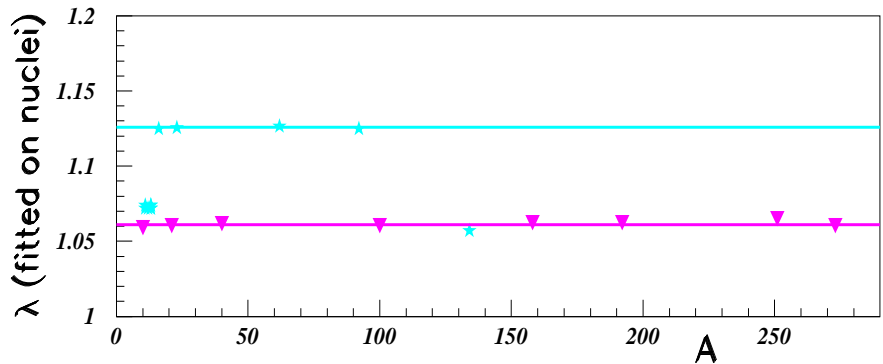}}
\caption{Variation of $\lambda$ values  versus A. Full purple triangles  show the values extracted from nuclei mass variations. Full sky blue stars show the values from nuclei energy level masses.}
\end{center}
\end{figure}

Several marks are not apparent, since they have the same values.
We observe two well separated $\lambda$ solutions, of mean values (horizontal lines)  1.061 and 1.126. Refering to equation (2.4), we observe that we have the same distribution, if $\lambda_2$ (n = 2)  = $\lambda_1^{2}$ (n = 1). This is exactly the relation between both experimentally extracted  values of $\lambda$. Such property was reported in \cite{sornette}, namely that  "DSI obeys scale invariance for specific choices of $\lambda$, which form an infinite set of values that can be written as $\lambda_{n}$ = $\lambda^{n}$. However, here,  the previous relation does not apply to successive sets of parameters needed to describe the various parts of a given DIS distribution, but the two different values are found when different distributions are considered.
\subsection{Overlapping fractals}
\begin{figure}[ht]
\begin{center}
\scalebox{1.3}[1]{
\includegraphics[bb=35 138  550 550,clip,scale=0.6]{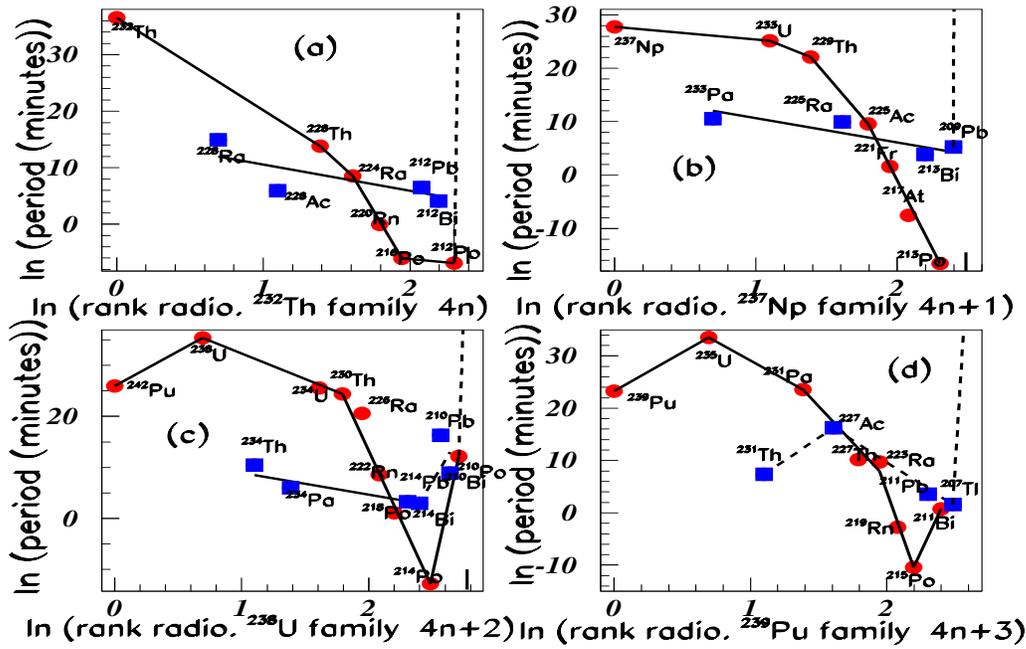}}
\caption{Log-log distributions of the four radioactive family periods. Full red circles correspond to $\alpha$ disintegrations; full blue squares correspond to $\beta^{-}$ disintegrations.} 
\end{center}
\end{figure}
Two examples of overlapping fractals are presented. The first one concerns the radioactive families, the second is observed in the figure showing an example of separation energies.

Fig.~37 shows the log-log distribution of the period (in minutes) of the four radioactive families. Insert (a), (b), (c), and (d) show respectively the results for the $^{232}$Th family (4n), 
the $^{237}$Np family (4n + 1), the $^{238}$U family (4n + 2), and the $^{239}$Pu family (4n + 3). The full red circles show the log of the period (in minutes) corresponding to the $\alpha$ disintegrations; the full blue squares correspond to the $\beta^{-}$ disintegrations. We observe overlapping fractals, with however often poor alignements, mainly those which correspond to $\beta^{-}$ disintegrations.

Fig.~38 shows the log-log distributions of the one nucleon and two nucleon separation energies from the mass number A = 146 nuclei.
\begin{figure}[ht]
\begin{center}
\scalebox{1.}[0.9]{
\includegraphics[bb=16 139 521 547,clip,scale=0.7]{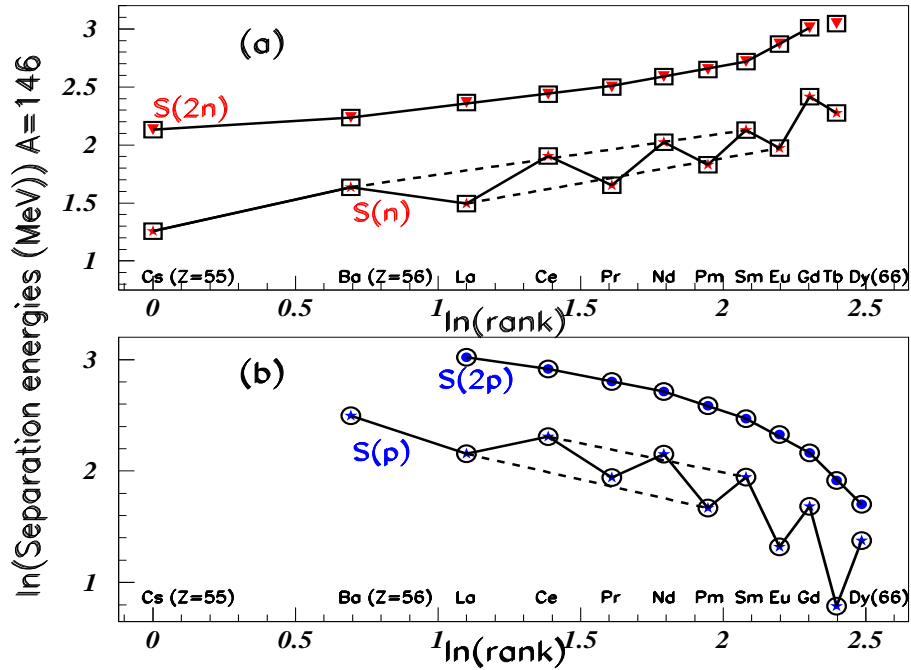}}
\caption{Log-log distributions of the separation energies from the mass number A = 146 nuclei (see text).} 
\end{center}
\end{figure}
Insert (a) shows the log-log distributions of the separation energies \cite{wapstra} of one neutron (full red triangles, encircled by black squares) and two neutrons (full red stars, encircled by black squares). Insert (b) shows for the mass number A = 146, the log-log distributions of the separation energies \cite{wapstra} of the one proton (full blue stars, encircled by black circles) and two protons (full blue circled encircled by black circles).  When the two nucleon separation energies display a "DIS" (four straight line segments), this is not the case for the one nucleon separation energies. These last distributions exhibit a stair case distribution induced by pairing effect. The one nucleon separation energies for the even N and Z 
nuclei are larger than the one nucleon separation energies for the odd N and Z 
nuclei. They, both, exhibit overlapping fractal property, namely alignement in the log-log distribution indicated by dashed lines. 
\section{General discussion on the parameter values}
Figure~39 shows the Im($\alpha$) versus Re($\alpha$) plot. Here we not only use the $\alpha$ values obtained from the fits performed above through the study of the hadron fractal properties, but also the parameters extracted from the nuclei masses and nuclei energy levels fractal properties.  We use also the $\alpha$ values \cite{boris} from the  study of the quark mass ratios. A great number of marks are not apparent, since they have the same values.
\begin{figure}[ht]
\begin{center}
\hspace*{-3.mm}
\scalebox{1.1}[1.5]{
\includegraphics[bb=4 286 520 520,clip,scale=0.5]{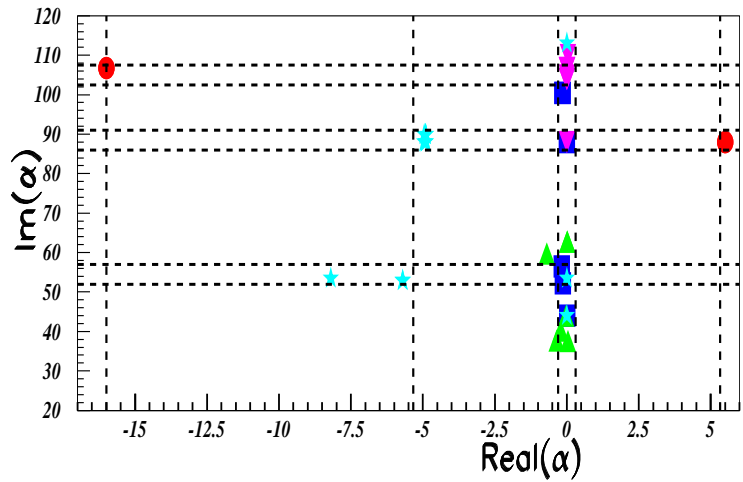}}
\caption{Plot of Im($\alpha$) versus Re($\alpha$). Full red circles show the $\alpha$ values which fit the elementary particle quark and lepton mass distributions; full up side green triangles show the same for meson data; full blue squares show the same for baryon data; full down side purple triangles show the same for nuclei yrast mass variation data; and full blue sky stars show the same for nuclei energy level mass data.}
\end{center}
\end{figure}

Full red circles show the $\alpha$ values which fit the elementary particle quark and lepton data; their values differ strongly from the others.

Full green up side triangles show the same for meson data;
the corresponding Re($\alpha$) is stable, close to zero, when the imaginary part varies from 37 up to 63.

Full blue squares show the same for baryon data; the real part of $\alpha$ is stable close to zero, when the imaginary part moves from 44 up to 101.

Full purple down side triangles show the same for nuclei mass variation data; here Re($\alpha$)~$\approx$0 is again stable and the imaginary part 
varies a little around  2*$\pi$/ln(${\bar \lambda_{1}}$) = 104.

Full blue sky stars show the same for nuclei energy level mass data. 
Many marks are not apparent, since they have the same value: $\alpha$~=~-4.9~+~89*i. The other six marks scatter, but  three of them have again a real part close to zero, and the real part of two of them, equals to -5.7 not very far from -4.9. 

The marks are not distributed randomly; this is emphasized by dashed lines. There are two distinct Im~($\alpha$) values for excited nuclei energy level masses. At the same values we observe  Im~($\alpha$) for mesons, quarks, leptons, and data from nuclei mass variation data. All data, except quark, lepton, and data from  nuclei energy level masses have Re~($\alpha$)~$\approx$~0.

In summary, the analysis of fractal properties of elementary particles masses, hadronic masses,  nuclei masses and nuclei energy levels, is performed with a rather small number of parameters, much smaller than the very large number of data analysed here.

\section{Conclusion}
\begin{figure}[ht]
\begin{center}
\scalebox{1.0}[1.0]{
\includegraphics[bb=47 325 521 520,clip,scale=0.8]{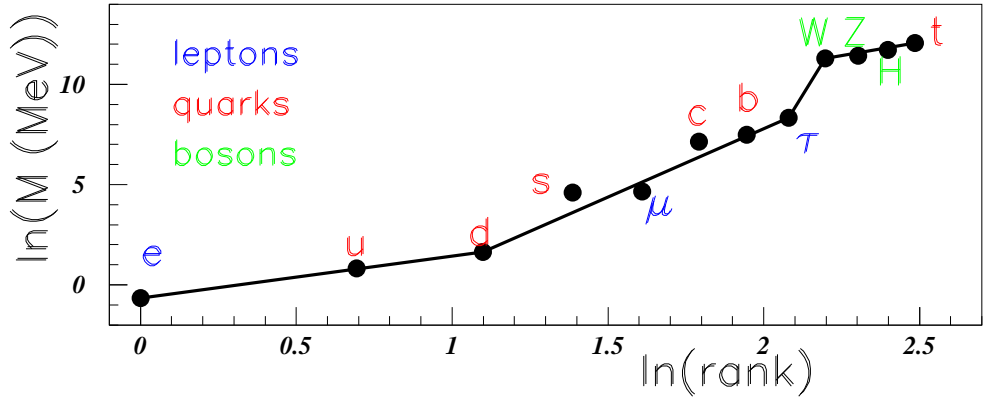}}
\caption{Log-log distributions of the fundamental particle masses (see text).} 
\end{center}
\end{figure}

Several conclusions can be drawn out, namely:\\
 - We have observed that the fractal properties with discrete scale invariance are observed in  fundamental  particle masses, as well as in hadronic and nuclei  masses.\\ 
 - The scale invariance is discrete since Im($\alpha$) is usually more than 18 times larger than Re($\alpha$).\\
 - We have observed and fitted, using equation (2.4), the oscillations of the successive  mass ratio distributions up to rank~$\approx$~10  for all mesons and baryon species. The various parameters included in the fits of the mass ratio distributions, depend little on the studied species.\\
 - This new symmetry allows to make several mass predictions. Indeed
 the fractal properties were used to predict some still unknown hadronic and nuclear masses.  They were also tentatively used to help to determine some unknown particle spins and unknown nuclei excited level spins \cite{bor4}, although sometimes without firm results.\\
  - We have observed a common framework between meson and baryon masses.\\
 -  We have observed a novel property between baryonic masses: the ratios between various families is constant.  In other words we observe, between different baryonic families, a constant ratio versus the rank, except for the first (sometimes the two first) rank(s). \\
  - This last property is not observed for meson families, and the difference between the baryon and meson behaviours remains to be understood.\\
 - The agreement with  fractal properties, is not so good
for the spectra of nuclei energy levels. However the comparison between the excited level masses of all baryonic species, display noteworthy properties allowing again a tentative
 prediction of several unobserved masses.   
  
  These observations can be generalized to different observables. Thus fig.~40 shows the log-log distribution of leptons, quarks, and bosons introduced in the increasing masses. Here all fundamental particle masses are introduced, except the neutrino masses. Indeed the neutrino masses are very low and unknown. The likelihood of their introduction inside an extension of fig.~40 with an unique alignement is unprobable, unless several sterile neutrinos would be found. However the three neutrinos can built a new straight line segment, which could be extended to some sterile neutrinos. It was noticed by Dolgov \cite{dolgov} that warm sterile neutrinos with a mass in the kev range are the most popular candidate for the sake of dark matter particles.
  
  The alignement in fig.~40 is noteworthy, except for the strange quark which mass may be too large. The same need for much smaller strange quark mass value was already suggested \cite{kartavtsev}. Fig.~40 shows that there is a common property between the masses of all fundamental particles: quarks, leptons and gauge boson masses.
  
   The same property was already found \cite{btib} when several relations between elementary particle masses were given, using only known physical constants without any arbitrary number. The quark-lepton similarity was also noticed \cite{hwang}.
\begin{figure}[ht]
\begin{center}
\scalebox{1.}[1]{
\includegraphics[bb=19 323 536 545,clip,scale=0.7]{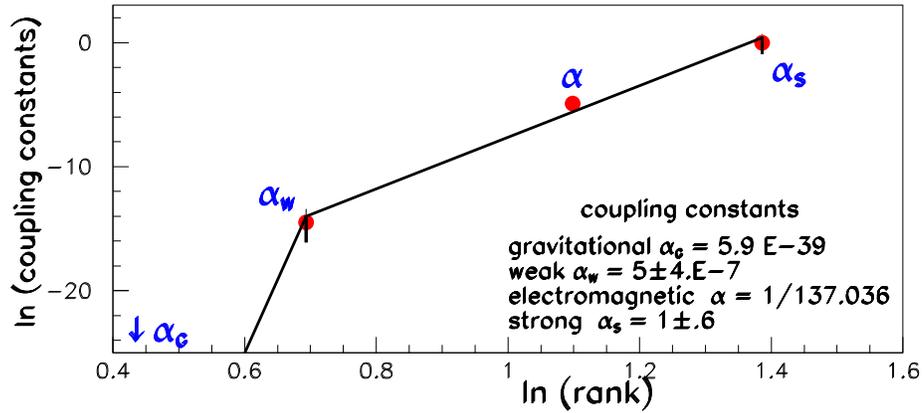}}
\caption{Log-log distribution of the four coupling constants: gravitational $\approx$ 5.9 10$^{-39}$, weak $\alpha_{W}$ = 5$\pm4 10^{-7}$, electromagnetic $\alpha$ = 1/137.036, and strong $\alpha_{S}$ = 1$\pm$ 0.6. } 
\end{center}
\end{figure}

  The fractal property allows eventually to add new masses between the "b" quark mass and the "W" boson mass. Indeed such introduction would reduce the slope between both particles in fig.~40. In that case the first new masses should be close to M = 9~GeV and 19~GeV. The same alignement between the "W" and "t" masses allows to predict the next mass could be close to M $\approx$ 210~GeV. 

Fig.~41 shows the log-log distribution of the four known coupling constants. Whereas the strong ($\alpha_{S}$ = 1$\pm$0.6), the electromagnetic ($\alpha$ = 1/137.036), and the weak ($\alpha_{W}$ = 5 $\pm$ 4 10$^{-07}$) coupling constants are aligned, this is not true for the gravitational ($\alpha_{G}$ = 5.9 10$^{-39)}$ coupling constant which is very low. Here also, there is room for tentatively new interactions between the gravitational and the weak interactions, giving rise to a new straight line segment or the extension of the one defined by the three largest couplings. In that case, the new couplings should be close to 
1 10$^{-25}$, 1 10$^{-17}$, and 1 10$^{-11}$.                                                                                                                                                              
However we have seen that an unique straight line is no more likely that straight line segments.

Other definitions for the gravitationnal coupling constant exist. If $\alpha_{G}$ = 1.752 10$^{-45}$, then the three new coupling constants could tentatively be:
g1=5 10$^{-29}$,  g2=1 10$^{-19}$,  and g3=5 10$^{-13}$. 

 An upper limit on the strength of an axial coupling constant for a new light spin one boson was reported recently \cite{piegsa}.
 
 It is worth saying that the tentatively suggested new particles or couplings are not a claim, but only interesting remarks.
\begin{figure}[ht]
\begin{center}
\scalebox{1.}[1]{
\includegraphics[bb=40 333 525 525,clip,scale=0.7]{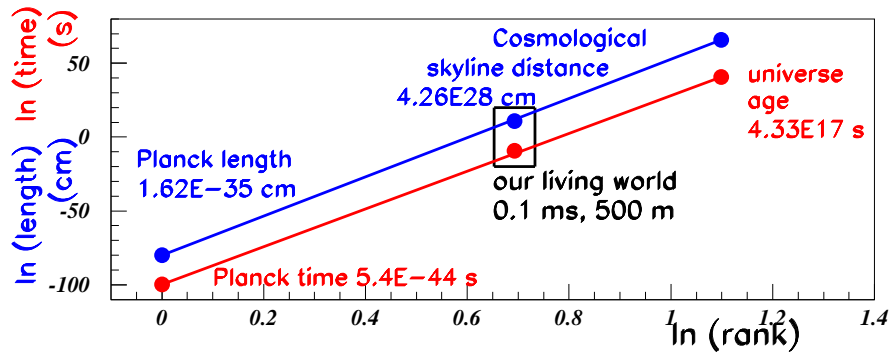}}
\caption{Log-log distribution of the minimal and maximal meaningfull values of time and distance.} 
\end{center}
\end{figure}

Fig.~42 shows the log-log distribution of the smallest meaningfull time (Planck time = 5.4 10$^{-44}$ s) and distance (Planck length = 1.62 10$^{-35}$ cm) and the largest meaningfull same values (universe age = 4.33 10$^{+17}$ s, and distance = Cosmological skyline distance = 4.26 10$^{+28}$ cm). These values are arbitrarily put at ranks 1 and 3. Then, inside fractals, our living world takes place in rank 2, exactly between both limits, namely at t = 0.1 ms and l = 500 m.

\section{Acknowledgments}
 I thank  Ivan Brissaud  for stimulating remarks and interest. I thank the organizers for the invitation and for warm hospitality during the conference.

\end{document}